\documentclass[12pt,preprint]{aastex}
\begin{document}
\title{Spectroscopic Analysis of Two Carbon Rich Post-AGB Stars}
\author{B.E. Reddy and David L. Lambert}
\affil{Department of Astronomy, University of Texas, Austin, TX 78712;ereddy,dll@astro.as.utexas.edu}

\author{G. Gonzalez}
\affil{Department of Astronomy, University of Washington, Seattle, Washington 98195;gonzalez@astro.washington.edu}

\author{David Yong }
\affil{Department of Astronomy, University of Texas, Austin, TX 78712;tofu@astro.as.utexas.edu}

\begin{abstract}
The chemical compositions of the C-rich pAGB stars IRAS 05113+1347
and IRAS 22272+5424 are determined from high-resolution optical
spectra using standard LTE model atmosphere-based techniques.
The stars are C, N, and $s$-process enriched suggesting efficient
operation of the third-dredge up in the AGB star
following a first dredge-up that
increased the N abundance. Lithium is present with an abundance
requiring Li manufacture. With this
pair, abundance analyses are now available for 11 C-rich pAGBs. A common history
is indicated  and, in particular, the $s$-abundances, especially the
relative abundances of light to heavy $s$-process elements, follow
recent  predictions for the  third dredge-up in AGB stars.
\end{abstract}


\section{Introduction}
The post-asymptotic giant branch (pAGB) stars, as the immediate successors of AGB
stars,  
can be used to probe nucleosynthesis and mixing processes in AGB stars. The spectra of pAGB stars
are simple relative to spectra of AGB and  planetary nebulae (PNe). In the case of AGB stars,
the presence of numerous molecular lines in their spectra makes
it difficult to derive accurate abundances. The central stars of PNe
to which pAGBs evolve are so hot that many 
elements do not provide lines in optical spectra. General properties of pAGBs
are well reviewed by Kwok (1993) and recently by Hrivnak (1997).

In the most advanced AGB stars, the predicted changes of surface composition
arise from the (third) dredge-up of material from the He-shell below the
H-rich envelope. Principal products added to the envelope and
atmosphere are $^{12}$C, and heavy elements synthesised by the
$s$-process involving neutron capture on a long timescale. Cool
carbon stars, i.e., stars with a carbon-rich (C/O $>$ 1) atmosphere,
are Nature's counterparts to the theoreticians' advanced AGB stars.
Carbon-rich pAGB stars have been identified, and this paper presents
abundance analyses of two such stars. Many pAGB stars are {\it not}
C-rich, and many, if not all, of the O-rich pAGB stars are also
not enhanced in $s$-process products. Examples include the RV Tauri
variables (Giridhar, Lambert, \& Gonzalez 2000) at the cool
end of the pAGB range, and high galactic latitude B-type supergiants, e.g., LSIV -12$^{o}$111
(Conlon et al. 1993) towards the hot end of
the pAGB range. Evidently, many stars depart the AGB before
the third dredge-up has transformed the O-rich giant to a C-rich
star.

The number of C-rich pAGB stars for which detailed abundance analyses are
available is small: 9 stars in total by Reddy et al. (1997, 1999), and
Van Winckel \& Reyniers (2000). Their compositions do resemble the
predictions for advanced AGB stars, but a large sample of stars is
needed to test thoroughly the theoretical predictions. The two pAGB
stars - IRAS 05113+1347 and IRAS 22272+5435 - analysed here are
classified as 
C-rich because their near- and far-infrared
spectra show emission features
at 3.3~$\mu$m, 3.4~$\mu$m,
and 21~$\mu$m attributed to carbon-bearing
PAH features (Kwok, Volk, $\&$ Hrivnak 1989;
Hrivnak, Kwok, $\&$ Geballe 1994). Basic data for the two pAGB stars are given in Table~1.
In this paper, we present abundance results of these two stars. 
Our analysis of IRAS~05113+1347 is the first to be published. IRAS~22272+5435
was analyzed previously by Za\v{c}s et al. (1995). Our results differ significantly from theirs, as we discuss below. 

\section{Observations}
Spectra of IRAS~22272+5435 and IRAS~05113+1347 were obtained with the 2.7~m
McDonald Observatory and the coud\'{e} cross-dispersed echelle spectrograph
(Tull et al. 1995) on June 14, 2000 and March 26, 2001, respectively.
Spectra have a resolving power of R$\approx$55,000 
as measured from FWHM of Th emission lines, and
S/N ratios of 80 - 400 at 6500~\AA.
The spectra run from 4000~\AA\ in the blue to 8800~\AA\ in the red with
gaps in the orders beyond 5600~\AA.
A high-resolution (R$\approx$45,000) spectrum for IRAS~05113+1347 was 
also obtained on December 31, 1999 with
3.5~m Apache Point Observatory telescope and an echelle spectrograph.
The spectrum runs from 4000~\AA\ in the blue to 10000~\AA\ in the red
with overlapping echelle orders.

Spectra have been reduced by standard procedures using the spectral reduction
package called IRAF\footnote{IRAF is distributed by the National
Optical Astronomical Observatories, which is operated by the
Association for Universities for Research in Astronomy, Inc., under
contract to the National Science Foundation.}. Telluric lines in the spectrum of IRAS~05113+1347
have been removed by dividing with a spectrum of a rapidly rotating hot star.
Selected spectral regions at 6785\AA\  and 7965\AA\ of both 
pAGB stars are displayed in Figures~1 and 2. Atomic
spectra are strikingly similar in both stars 
(Fig~1). Both stars show
strong $s$-process lines.
The metallic lines are weak 
suggesting metal deficiency. IRAS~05113+1347 shows photospheric 
CN lines (Fig~2). Absence of photospheric
CN lines in the spectrum of IRAS~22272+5435 may be due to its higher
$T_{\rm {eff}}$.

\section{Analysis}

\subsection{Selection of lines and $gf$ values}
Selection of clean lines with reliable $gf$-values 
is very important in reducing uncertainties in the abundances.
In selecting lines we made use of
the very high resolution spectrum of the cool giant Arcturus 
(Hinkle et al. 2000), as well as the
solar spectrum (Wallace, Hinkle $\&$ Livingston 1993). To avoid overcrowding of lines
and uncertainties in continuum placement, we examined
the green and red parts but not the blue part of the spectra. 
We limited the analysis to
lines of moderate strength, equivalent width (W$_{\lambda}$)~$\leq$ 200~m\AA). Our line list comprised 
about 150 atomic lines belonging to 25 elements.

Choosing accurate $gf$-values is important in the reduction of
errors.
We concentrated on lines which have well determined $gf$-values
experimentally or theoretically. The $gf$-values for most of the Fe\,{\sc i} and Fe\,{\sc ii} lines come from
laboratory measurements of either the Oxford or Kiel-Hanover groups; 
Lambert et al. (1996) discussed the $gf$-value
measurements. For other elements, 
we chose accurate  $gf$-values from a variety of sources:
Cr (Blackwell et al. 1986a), Si (Garz 1973), Ti\,{\sc i} (Blackwell et al. 1986b), 
V (Whaling et al. 1985),
Na, Al (Lambert $\&$ Warner 1968), Ca (Smith $\&$ Raggett 1981), Y\,{\sc ii} ( Hannaford et al. 1982)
and for C, N, and O (Wiese et al. 1996). Li, Sc, and Mn exhibit  hyperfine structure (HFS).
For Sc and Mn, we adopted HFS data of Kurucz (2000), as discussed by Prochaska \& McWilliam
(2000). For Li we adopted the HFS and isotopic structure given by Andersen et al. (1984). For the
rest of the lines, the $gf$-values come either from R.E. Luck's compilation (private communication)
or Kurucz's (2000) atomic line list. To check the reliability of the $gf$-values
we derived abundances for the solar spectrum using solar equivalent widths measured
either by Meylan et al. (1993) or from our solar spectrum
at R$\approx$60,000 obtained from the McDonald Observatory
2.7~m telescope. We used the Kurucz (1995) convective solar model
with a microturbulence of 1.15 km s$^{-1}$. We accepted all the individual $gf$-values which yield
abundances within $\pm$0.2~dex of mean solar abundances (Grevesse $\&$ Sauval 1998) 
determined using the lines having reliable laboratory $gf$-values.
 
\subsection{Atmospheric Model Parameters}
In this study, we adopted plane parallel, and line blanketed LTE model atmospheres computed
by Kurucz (1995) using the ATLAS9 code and
extracted from http://cfaku5.harvard.edu.
Final models were found by interpolation. 
We assume that the standard helium abundance is appropriate to these pAGB stars.
A good discussion of the Kurucz 
model atmospheres is given by Castelli et al. (1997).

Initial values of model parameters
$T_{\rm eff}$ and log~$g$ were estimated 
from their spectral classification and photometry. Hrivnak (1995) classified
IRAS~05113+1347 as G8Ia and IRAS~22272+5435 as G5Ia 
using low-resolution spectra.
The spectral types of G8 and G5 translate into $T_{\rm eff}$ of 4260~K and 4460~K, 
respectively (B\"{o}hm-Vitense 1972).
We estimate log~$g$$~\approx$~1.0 from the luminosity class Ia and log $g$ tables. The values of $T_{\rm eff}$
can also be estimated from intrinsic B-V colors. The interstellar extinction
values in the direction of the stars were taken from Hrivnak (1995). 
The intrinsic B-V colors for supergiants (B\"{o}hm-Vitense 1972)
yield 5100~K for both stars. 

Using these initial estimates, 
we derived more accurate model parameters: $T_{\rm eff}$, log~$g$, 
microturbulent velocity ($\xi_{t}$), and metallicity
for the program stars from the high-resolution spectra. The values
of $T_{\rm{eff}}$ are derived from the excitation of Fe\,{\sc i} lines. 
We chose 23 Fe\,{\sc i} lines with lower excitation potentials ranging from
0.85~eV to 4.6~eV. For a given log~$g$ =1.0~dex and an assumed $\xi_{t}$ = 5.0 km s$^{-1}$, 
the values of $T_{\rm{eff}}$ are determined from the requirement
that the
individual Fe\,{\sc i} abundances
be independent of excitation potentials (Fig~3a, 3b). 
The values of log~$g$ are determined from the
requirement  that 
neutral and ionized iron lines yield similar abundance for the chosen $T_{\rm eff}$.
The values of $\xi_{t}$ are derived by forcing
a zero slope on the abundances of lines Fe\,{\sc i}, Fe\,{\sc ii}, and Ni\,{\sc i} as a function of 
W$_{\lambda}$. This method was iteratively repeated 
until we obtained a self-consistent set of parameters. The adopted
model parameters are given in Table~2.

\subsection{Radial Velocities}
Radial velocity 
measurements
not only provide information on the systemic velocity, but also
yield information on the atmospheric structure: velocity gradients in the
atmosphere, dust shells around the stars etc.
Radial velocities of various atomic and molecular absorption
lines are given in Table~3. With the exception of resonance lines, the
atomic lines are of photospheric origin and provide the systemic velocity:
the mean heliocentric velocities are $V_{r}$ = 7.6~$\pm$ 1.5~km s$^{-1}$ for
IRAS~05113+1347 and $-$42.4~$\pm$1.0~km s$^{-1}$ for IRAS~22272+5435. 
The velocity of IRAS~22272+5435 is consistent given the probability of a
small amplitude variation with the value $V_{r}$ = $-$39.7 $\pm$ 0.5~km~s$^{-1}$ also
measured using optical lines (Za\v{c}s et al. 1995). Observations of CO 
millimeter emission lines also provide
an estimate of the systemic velocity provided that the emitting gas is
uniformly distributed around a star. These estimates match our velocities:
$V_{r}$ = 7.2 km~s$^{-1}$ (Hrivnak $\&$ Kwok 1999), and $-$41.0 $\pm$ 1.0 km s$^{-1}$
(Loup et al. 1993) for IRAS~05113+1347 and IRAS~22272+5435, respectively.

Circumstellar C$_{2}$ and CN lines are seen in the spectra (Fig~4). 
In the case of IRAS~05113+1347,
Photospheric
CN but not C$_{2}$ lines are also seen but at the systemic velocity, V$_{r}$. 
For IRAS~05113+1347,
Bakker et al. (1997) reported the detection of C$_{2}$ Swan and Phillips, and
CN Red system lines from several vibrational bands. Many rotational
lines of the C$_{2}$ bands, as expected for a homonuclear molecule, are seen
for each band. The circumstellar lines of the CN bands are restricted to
low-$J$ lines. Our measurements of C$_{2}$ Swan 0-2 and 0-3 and CN Red system 1-0 lines
shows that they are shifted by $-$10.2 $\pm$ 1.0~km s$^{-1}$ relative to the
systemic velocity, a velocity difference identified as the expansion velocity (V$_{exp}$) of that
region of the circumstellar shell containing the C$_{2}$ and CN molecules. This expansion
velocity is similar to the estimate of 13.1 km s$^{-1}$ from the width
of the CO millimeter emission lines (Hrivnak $\&$ Kwok 1999). 

Our detections
of optical circumstellar molecular blue-shifted lines for IRAS~22272+5435 are the first
for this star which is too warm to show photospheric molecular lines. Swan system lines give an expansion
velocity of 6.3 $\pm$ 0.3 km s$^{-1}$ which compares with 8.0 to 12 km s$^{-1}$ from
the widths of CO millimeter lines (Loup et al. 1993). 

\subsubsection{Profiles of Alkali Resonance Lines} 
An exciting observation is the apparent presence of the Li\,{\sc i} 6707~\AA\
resonance doublet weakly in absorption in both spectra. But excitement is tempered by the
additional observation that the line is displaced redward by about 0.2~\AA\ from
the wavelength expected of a photospheric line. Understanding this shift and the
line profile is a prerequisite to determining the stellar lithium abundance.

A clue to the correct interpretation of the Li\,{\sc i} line is, in the case
of IRAS~05113+1347 provided by inspection of the resonance line profiles of
the more abundant alkali atoms Na and K. Figure 5a shows the profiles
of the Na\,{\sc i} D$_{1}$, the K\,{\sc i} 7699\AA, and the Li\,{\sc i} doublet
as observed on 1999 December 31. Emission in the core of the Na\,{\sc i}
D lines exceeds the local continuum. This emission, which is from the star and not the nightsky,
divides the absorption profile into two components which we label A and B. Similar
emission may be imagined in the K\,{\sc i} line but it is
evidently weaker with the line profile more strongly suggesting the presence
of the two absorption components. The peak Na\,{\sc i} emission is at a velocity of
7 km s$^{-1}$, which is the photospheric/systemic velocity (Table 3). This suggests
that emission may arise from a large shell, presumably a part of the expanding circumstellar
envelope responsible for the CO millimeter emission. 
The spectrum from March 26, 2001 does not show Na D emission
but the components A and B remain. Profiles of the Li\, {\sc i} and K\,{\sc i} resonance lines are
almost unchanged between the two observations. Emission in the Na D lines, if present in IRAS~22272+5435,
is very weak. The absorption for this star consists of at least four components: A and B with the  
same presumed origins as the A and B components in IRAS~05113+1347, and two additional
components assigned an interstellar origin and labelled IS in Figure 5b. 

For both stars, absorption component A is 
redshifted relative to the systemic velocity by about 10 km s$^{-1}$.
Similar velocity shifts were found for the two pAGB stars studied
previously (Reddy et al. 1999). That the component is similar for
four stars suggests that it has a common origin, and is not, for example, simply an
intruding interstellar line. Two possibilities may be considered: the component is primarily of
photospheric origin but redshifted because the line is preferentially formed in
cool downdrafts; the line is formed in infalling circumstellar gas. Present evidence suggests
the latter is more probable because it appears that only resonance lines of
low ionization potential atoms are affected. If the former explanation were
valid, one would expect to see a velocity shift between low excitation lines of Na\,{\sc i}, and Ca\,{\sc i}
(for example) and high excitation lines of C\,{\sc i} and O\,{\sc i} but Table~3 shows no differences. Of course,
the presence of infalling (component A) with outflowing (component B) implies a complicated circumstellar
shell but such a juxtaposition is not unknown for giant stars.

If component A is taken to be the photospheric line shifted by convective downflow, central
emission is seen in all observations of the Na D and K\,{\sc i} profiles. The predicted
photospheric K\,{\sc i} profile (Fig. 5) is based on the inferred normal 
abundances for [Fe/H] $\simeq$ $-$0.8, and
the Kurucz model. The true profiles in the actual convective photospheres are surely somewhat
different. The emission with component B constitutes a P Cygni profile provided by circumstellar alkali atoms.
Lithium may be similarly affected.

\subsection{Abundances}
Elemental abundances were determined using  the code MOOG     
(Sneden 1973) in combination with the adopted model atmosphere.
Most of the
selected lines are unblended and 
computing abundances from equivalent width measurements works well. For a few elements
which suffer from blends and/or 
hyperfine structure, abundances were determined using spectrum synthesis.

A summary of abundances for IRAS~05113+1347 and IRAS~22272+5435 is given in Table~4
where $\it{n}$ is the number of lines, and $\sigma$ is
the standard deviation representing
scatter among lines. 
The quantity log~$\epsilon$(X) = log ($N_{\rm {X}}/N_{\rm {H}}$)+12 is
the abundance of an element X relative to a hydrogen abundance of 12
on a logarithmic scale. The ratio [X/H] = log~$\epsilon (\rm {X}_{\star}$)$-$ log~$\epsilon (\rm {X}_{\odot})$ 
refers to the stellar
abundance relative to the solar photospheric abundances of Grevesse $\&$ Sauval (1998). 
The abundances relative to iron [X/Fe] = [X/H]$-$[Fe/H] are computed  using our derived
stellar iron abundances of [Fe/H].

Errors in the abundance analysis come from different
sources. Errors in the adopted $gf$-values and in the measured $W_{\rm {\lambda}}$
introduce uncertainties. These errors can
be represented by the standard deviation, $\sigma$ (Table~4)
that measures the scatter among the individual lines. The true error, i.e.,
standard deviation of the mean $\sigma/\sqrt{n}$, is less for elements
represented by many lines. For abundances which are derived by
fewer than 3 lines the real error may be large, and we adopted
$\sigma$ =0.2~dex for these sources of error.

The abundances are also
model dependent. The four parameters: $T_{\rm{eff}}$, log~$g$, $\xi_{\rm {t}}$, and
[M/H] that are required to select the atmospheric model are somewhat coupled.
The uncertainty in the $T_{\rm {eff}}$ can be estimated by examining the
relation between the lower excitation potential and abundance. The slope of the fit to the
observations is quite sensitive to $T_{\rm {eff}}$.
The error (standard deviation) in the slope 
$\sim$ 0.03 translates into an uncertainty in the $T_{\rm {eff}}$ of 150~K.
Similarly, we estimated uncertainty in $\xi_{\rm {t}}$. In the case of log $g$,
the error is estimated by introducing 0.1~dex difference in abundance between neutral and ionized lines.
Thus, the estimated
uncertainties in the input model parameters are as follows:
$\delta$$T_{\rm{\rm eff}}$ = 150~K, $\delta$log $g$ = 0.50~dex,
$\delta$$\xi_{\rm {t}}$ = 0.50 km s$^{-1}$, and $\delta$[M/H] = 0.25~dex. 
The uncertainties in the
abundance ratios as a result
of uncertainties in the model
parameters are given in the Table~5 for IRAS~05113+1345. The values in the Table~5 show that
the abundance ratios are less affected by uncertainties in the [M/H] and $\xi_{\rm {t}}$.
The uncertainty in the gravity of 0.5~dex
changes the ratios up to 0.15~dex. The 
abundance ratios determined from the lines of
high excitation are found to be sensitive to changes in $T_{\rm{eff}}$ (Table~5). 
For each element the quadratic sum of the uncertainties in the four model parameters
is given as $\sigma_{m}$.

Our results for IRAS~22272+5435 differ quite considerably in several key respects from
those published by Za\v{c}s et al. (1995). 
The differences are not related to slight differences in the adopted fundamental parameters.
There is a small difference in the derived
iron abundances: we obtain [Fe/H] = $-$0.8 but Za\v{c}s et al. got [Fe/H] = $-$0.5.
Large differences are found for many other elements. For example, they reported several cases
of extraordinary enhancements relative to iron: e.g., ratios for elements around
iron include [Cr/Fe] = $+$2.0, [Ni/Fe] = $+$1.5, and [Zn/Fe] = $+$2.2. In all
cases, our ratios are close to the expected values for a moderately metal-poor star,
that is [Cr/Fe] = [Ni/Fe] = [Zn/Fe] $\simeq$ 0.0. Although our analyses agree that
the $s$-process elements are severely overabundant, abundances in the two analyses differ
by up to 1.0 dex. These differences may perhaps be traceable to the lower resolution
of the spectra used by Za\v{c}s et al., and their tendency to include strong lines
in the abundance analysis. Scrutiny of their list of measured lines showed several
cases where we identified much weaker lines with the transition that they used.

\subsubsection{Carbon, Nitrogen, and Oxygen}
The abundance of carbon is determined using three different indicators: permitted C\,{\sc i} lines,
the forbidden C\,{\sc i} line at 8727.1~\AA\ (this line
is not measured in the spectrum of IRAS~22272+5435),
and lines of the CN red system at 8035~\AA. 
The
forbidden C\,{\sc i} line is a superior C abundance indicator as it is 
unaffected by NLTE (Gustafsson et al. 1999).
However, this line is blended with Fe\,{\sc i} and CN lines.
Fortunately,
the Fe\,{\sc i} and CN contributions to the blend at this metallicity are small.
All known blends are taken 
into account in the spectrum synthesis. For IRAS~05113+1347, the carbon abundance from this
forbidden line is 0.2~dex less than that derived
using the permitted C\,{\sc i} lines, but the error estimate
$\delta$$T_{eff}$ = 150~K
causes a difference of 0.19~dex in abundances derived from permitted and forbidden C\,{\sc i} lines. 
Non-LTE effects may also be a factor.
The CN Red system lines yield
a higher abundance by 0.3~dex than the forbidden line. This may be due to the uncertainty in the
abundance of N which is an input in the synthesis of CN lines and
is based on two weak N\,{\sc i} lines in the red.
Given the
uncertainties in the model parameters the derived C and N abundances 
from different abundance indicators are in good agreement. 

The abundance of oxygen is determined using both forbidden and permitted lines.
The oxygen abundance (log $\epsilon$([O\,{\sc i}])=8.37) determined from [O~I] lines 
at 6300.3~\AA\ and 6363.8~\AA\ for IRAS~05113+1347 is found to be in good agreement
with the abundance derived from the permitted O\,{\sc i} line at 9265.9\AA.
For IRAS~22272+5435, the oxygen abundance comes from the [O~I] line at 6300.3~\AA.
Final results of C, N, and O
abundances are given in Table~6. Abundances of C and O are the simple mean
of abundances derived using different indicators. Total sum of C, N, and O
abundances ($\Sigma{CNO}$) relative to Fe and ratios C/O and N/O are also given in Table~6.

\subsubsection{s-process elements}
A majority of
the s-process
lines used here come from Kurucz's (2000) line list.  
Many of these lines
do not exist in the solar
spectrum. 
Given the large number of lines in the Kurucz list
it is difficult to know that the $s$-process line is the dominant contributor. To
make sure, 
we computed equivalent widths for all the nearby lines. 
Molecular
CN and C$_{2}$ contributions to the lines were also taken into account.
This procedure allowed
us to identify quite a number of reasonably clean $s$-process lines.
The $s$-process lines, their atomic data, and abundances are given in the
Table~8. 

\subsubsection{ Spectrum Synthesis: Li Abundance}
Extraction of the Li abundance is important because lithium plays
an important role in understanding the evolutionary state of pAGB
stars.
 Here, the abundance analysis
is complicated by the fact that the dominant absorption attributable to
the Li\,{\sc i}
6707\AA\ line is at the velocity of the infalling material labelled
as component A. Component B, definitively circumstellar, may also
contribute absorption to confuse further a contribution from
 a Li\,{\sc i} photospheric line.
Finally, emission may fill-in the photospheric Li\,{\sc i}
absorption in the case of IRAS 05113+1347.
To within the errors of measurement, identical velocity shifts
of a line likely to be the
photospheric
Li\,{\sc i} line are found for this pair of pAGBs as well as
and the pAGBs IRAS 02229+6208 and IRAS 07430+1115 (Reddy et al. 1999).

Our syntheses of the Li region used atomic lines listed by Kurucz (2000),
and CN Red system lines (Cunha, Smith, \& Lambert 1995). The Arcturus spectrum
was synthesized and, as a result, minor modifications were made to the
adopted $gf$-values. We also included 
the $s$-process ionized line required
by Lambert, Smith, \& Heath (1993) to fit spectra of Ba\,{\sc ii} giants.
Selected syntheses are shown in Figure 6 where it is obvious that
the observed spectrum cannot be matched satisfactorily.
These syntheses leave unmatched the
absorption to the red that is
attributed to a Li\,{\sc i} line 
from the circumstellar shell (component A). The presence
of two Li\,{\sc i} components is not at all unprecedented having
been reported for a Li-rich (O-rich) giant in the globular cluster
NGC\,362 (Smith, Shetrone, \& Keane 1999), and for the Li-rich
semiregular variable, also O-rich, W LMi (Giridhar, Lambert, \& Gonzalez 2000).
Alternative interpretations seem implausible.
If photospheric lithium is assumed to be rich in $^6$Li,
a fair fit to the spectrum is
obtained, as noted previously by Reddy et al. (1999);
the $^7$Li - $^6$Li wavelength shift is about 7 km s$^{-1}$ or only
slightly less than
the redshift of component A, but the presence of $^6$Li is considered most
improbable. A  V\,{\sc i} line at 6708.07\AA\ might be
deemed responsible for component A but the $gf$-value needed to fit the
line in Arcturus must be increased by nearly 3 dex to match component A in the pAGBs.
Finally, it is unlikely that the component
is an unrecognized $s$-process line; no
such line was required by Lambert et al. (1993) in their syntheses of
spectra of Ba\,{\sc ii} giants including stars for which the Ce\,{\sc ii}
lines are of comparable strength to the lines in the
pAGB stars. (Technetium appears not to have a line at 6708.0\AA.)

If component A is assigned to circumstellar lithium atoms
and circumstellar emission is neglected, a fair assumption given that the
spectrum of IRAS~05113+1347 at 6707\AA\ is not significantly different when the Na D
emission was very prominent, the syntheses of the photospheric component give
log $\epsilon$(Li) $\simeq$ 1.8 for IRAS 05113+1347, and
$\simeq$ 2.1 for IRAS 22272+5435.

\subsubsection{Spectrum Synthesis: $^{12}C/^{13}C$ ratio }
To derive the carbon isotopic ratio, we selected lines of the CN Red system in the spectral region
8000-8010~\AA. This is the region where a strong blend of $^{13}$C features 
at 8000.4~\AA\ and clean $^{12}$CN lines are located. 
There are other $^{13}$CN features but most of them
are buried in stronger  $^{12}$CN lines. The molecular data:
line positions, excitation values, and oscillator strengths are taken from
the investigation of de Laverny $\&$ Gustafsson (1998; private communication 
from de Laverny). The list includes all the $^{12}$CN lines
given by Davis $\&$ Philips (1963)
and $^{13}$CN lines given by Wyller (1966).
Additional input data required to produce
synthetic spectrum is the dissociation energy that we adopted 7.75~eV (see Lambert 1993) for both the
molecules. The region 8000-8010~\AA\
was computed using the adopted line list for the Arcturus model atmospheres and the abundances.
The resultant spectrum was compared with the observed high-resolution spectrum of Arcturus
to test the accuracy of the molecular and atomic line data.
The predicted spectrum matches nicely the observed Arcturus spectrum and this also yields
$^{12}$C/$^{13}$C = 6.0 $\pm$ 1.0 which is in very good agreement with the isotopic ratio
measured for Arcturus (Griffin 1974).

We computed the spectrum for IRAS~05113+1347.
(Photospheric CN lines are not seen in the spectrum of
IRAS~22272+5435).
We fit the
$^{12}$CN for the derived N abundance by changing the C abundance.
For a C abundance of 8.90$\pm$0.1~dex the $^{12}$CN lines are well fitted (Fig~7).
The $^{13}$CN feature
blended with $^{12}$CN at 8002.0~\AA, the blends of pure $^{13}$CN line at 8004.5~\AA, and at 8006.0~\AA\
are  
sensitive to the changes in the isotopic ratio $^{12}$C/$^{13}$C. The spectrum was predicted for
three different carbon isotopic ratios and compared with the observed spectrum of IRAS~05113+1347.
The features at 8002.0~\AA\ and 8004.5~\AA\ indicate a lower limit of $^{12}$C/$^{13}$C$>$ 25.
A slightly smaller value is suggested by the $^{13}$CN feature at 8006.0. We
give greater weight to the feature at 8004.5~\AA\ as this is stronger and pure $^{13}$CN.
The limit $^{12}$C/$^{13}$C $>$ 25 is in good agreement with the typical 
lower limit of $^{12}$C/$^{13}$C $\geq$ 20 derived for pAGBs,
including IRAS~05113+1347,
by Bakker et al. (1997) using circumstellar molecular CN red lines.

\newpage
\section{Stellar Evolution and Chemical Composition}
Inspection of the compositions of the two pAGB stars (Table~4 \& 6) leads to four
broad conclusions, which we amplify below: (i) their composition is normal except
for those elements expected to be affected during evolution prior to the pAGB phase;
(ii) lithium is present at an abundance that probably implies lithium production in a
post main sequence phase; (iii) adjustments to initial C and N abundances occured as a
result of mixing with H- and He-burnt material; (iv) substantial enrichment with $s$-processed
material is traceable to mixing from the He-shell of their AGB progenitor. Collectively,
these conclusions suggest the stars have evolved from thermally-pulsing AGB stars.
The pair are compared with published results for 9 other C-rich pAGB stars (Table~7), which
support the four broad conclusions.

Even in the case of thermally-pulsing AGB stars, many elements are expected to retain their
initial abundances (relative to iron). This is the case here: Na, Al, Si, S, Ca, Ti, and the 
iron-group from V to Ni, and Zn show the ratio [X/Fe] reported for normal dwarfs and red giants
with [Fe/H] $\simeq$ $-$0.8. (Possibly, [Ca/Fe] $\simeq$ 0.0 is less than the expected
value of 0.2~dex 
as a result of Galactic chemical evolution.)
This demonstration of normal abundances serves to exclude the possibility that the
photospheres are affected by processes of dust-gas selective fractionation that are
seen to have affected some warmer pAGB stars (Waelkens et al. 1992), and the warmest RV\, Tauri variables
(Giridhar, Lambert, \& Gonzalez 1999). In particular, our pair of pAGB stars show
normal [S/Fe] and [Zn/Fe] ratios which are potent indicators of dust-gas fractionation.

\subsection{Carbon, Nitrogen, and Oxygen}
Carbon, nitrogen, and oxygen abundances are key tracers of the evolutionary and nucleosynthetic
history of red giants and their descendants. Low mass red giants are predicted to
experience three episodes in which the deep convective envelope brings nuclear-processed
material from the interior to the atmosphere. The first episode known as the first
dredge-up occurs when the star ascends the red giant branch, after exhaustion of hydrogen in the core
of the main sequence star. This brings mildly CN-cycled material to the surface and, therefore,
reduces the C abundance and enhances the N abundance with the constraint that the sum
of the C and N abundances is preserved (Iben 1964). Oxygen is predicted and observed to be
unaffected by the first dredge-up.

Additional changes of surface abundances occur through what is known
as the third dredge-up phase known to affect AGB stars. In this phase, material from the
He-shell around the C-O electron degenerate core is mixed into the base of the deep
convective envelope. The added material has been exposed to He-burning via the 3$\alpha$-process
and to $s$-processing. The principal product of the former is $^{12}$C with an admixture
of $^{16}$O. An AGB star is expected to experience the third dredge-up repeatedly
until it evolves to the pAGB phase and beyond. (Even then a final episode is possible.)
Through the third dredge-up, the O-rich giant may be converted to a
C-rich AGB star.
In its simplest theoretical implementation, the series of third dredge-up
events do not involve  exposure of the H-rich envelope to the H-burning
CN-cycle, that is the carbon added from the He-shell is not converted
to nitrogen. Very luminous AGB stars can experience H-burning at the base
of the H-rich envelope in the long intervals between times when the He-shell
is ignited. Such stars said to experience hot bottom burning (HBB),
convert carbon to nitrogen, reduce the $^{12}$C/$^{13}$C ratio, and
may be reconverted from a C-rich to an O-rich star.

In order to compare the observed C, N, and O abundances with theoretical
expectations, we begin with establishing the initial abundances for
these slightly metal-poor stars, and, then, we comment on the changes
expected from the first dredge-up which establish the expected surface
abundances for the AGB star prior to the onset of the third dredge-up.
Observed abundances for our two stars and eight others analysed
previously are summarized in Table 6.

Initial abundances for C and O are taken from Gustafsson et al. (1999)
who analysed the [C\,{\sc i}] 8727 \AA\ line in a sample of
main sequence stars, and collated results on the O abundance for a
similar and overlapping sample. Fits to the derived abundances gave

$[C/Fe] = (-0.17\pm0.03)\times [Fe/H] + (0.065\pm0.008)$

and

$[O/Fe] = (-0.36\pm0.02)\times [Fe/H] - (0.044\pm0.010)$

Measurements of the nitrogen abundance are less extensive. We adopt
[N/Fe] = 0 (Clegg, Lambert, and Tomkin 1981).

The inferred initial abundances are given in Table 6 along with the
observed values for the pAGB stars. Relative to initial values, the
stars are clearly enriched in carbon and nitrogen, but oxygen has retained
about its initial abundance. These are in line with our sketch of
the expected changes: the carbon enrichment is attributed to the
third dredge-up, and the nitrogen enrichment to the first dredge-up,
and, as expected, oxygen is almost unaffected by the dredge-ups.

From Table 6, the mean difference between the O abundance of the pAGB
stars and their inferred initial O abundance is 0.2 dex.
Given the possibility of systematic errors in comparing abundances
derived from main sequence and pAGB stars, this difference is compatible with
the expectation that oxygen is unaffected by evolution from the main sequence
to the pAGB stage. It may be noted that
a study of post first dredge-up red giants
with metallicities similar to those of stars in Table 6
concluded that oxygen was undepleted (Cottrell \& Sneden 1986).

There is evidence, primarily from the low observed and higher predicted
$^{12}$C/$^{13}$C ratios, that predicted  changes for carbon and nitrogen
brought by the first dredge-up are less severe than observed changes.
It is, however, expected that a red giant's nitrogen abundance will
approach the sum of the star's initial C and N abundances. Considering
this, we list in Table 7 this sum alongside the observed N abundances
for the pAGB stars. On average, the latter are equal to the
sum of the initial C and N abundances: the mean difference is less than
0.1 dex. This is certainly compatible with the idea that the nitrogen
abundance of AGB and pAGB stars was unaltered by the third dredge-up
and subsequent events.

Cottrell \& Sneden's analyses, however, do not fully support this
conclusion. Their carbon abundances show the expected underabundance,
for the first dredge-up, especially if the initial abundances are
taken from Gustafsson et al. (1999). Low $^{12}$C/$^{13}$C ratios
reported by the authors clearly show that a first dredge-up has
occurred. Oddly, nitrogen was reported to be underabundant: [N/Fe] =
$-$0.05 was found in the mean. This is surprising. We suppose that this
anomaly is either an artifact of the analysis and/or the assumption that
[N/Fe] = 0 is incorrect for the initial abundances.
If, of course, [N/Fe] $<0$ initially, our conclusion that the first
dredge-up accounts for the observed nitrogen abundances of pAGB stars
is invalid.

The first dredge-up reduces the surface $^{12}$C/$^{13}$C ratio.
Cottrell \& Sneden's measurements give a mean ratio of 16 for their
sample of mildly metal-poor giants. Available determinations for the
pAGBs including our lower limit for IRAS 05113+1347 give higher
values. These are consistent with addition of $^{12}$C to the envelope of
a red giant without exposure to the CN-cycle, which would convert some
$^{12}$C to $^{13}$C and $^{14}$N. Thus, the $^{12}$C/$^{13}$C ratios
and the nitrogen abundances support the idea that little of the
$^{12}$C from the third dredge-up has been processed by protons.

Carbon in pAGBs is considerably enhanced relative to its initial
abundance and certainly to its post first dredge-up abundance. The stars
are clearly C-rich, i.e., C/O $>$ 1. Table 6 shows that C/O $\simeq$ 3 for
IRAS 05113+1347, and 2 for IRAS 22272+5435. The sensitivities of the abundances
to the atmospheric parameters are so similar for the carbon and oxygen lines
that the derived C/O ratio cannot be plausibly decreased to less than unity.
We assume that fresh carbon was added through the third dredge-up events.

In summary, the C, N, and O abundances are compatible with the view that
these C-rich pAGB stars evolved directly from AGB stars that were
converted to carbon stars as a result of the third dredge-up. It remains
to examine if the $s$-process and lithium abundances may be integrated into
this picture.

\subsection{The $s$-process}
Both stars are severely enhanced in $s$-process elements (Table~7) which we identify as a
signature that the stars experienced an efficient third dredge-up in their AGB phase.
In this respect, our pair of pAGBs are similar to the other analyzed C-rich pAGBs.
Models indicate that low mass AGB stars run their $s$-processing in the He-shell with the
neutron source $^{13}$C($\alpha$, n)$^{16}$O. To provide sufficient
neutrons, $^{13}$C must be generated in the H-He intershell region by a slow
mixing of protons into the He-rich layers. Presently, the mass of $^{13}$C in the shell at the onset
of $s$-processing is not calculable from the first principles but must be specified
as a free parameter. 
This free parameter is a controlling influence on the neutron flux.
The higher the neutron flux the greater is the abundance
of $\it {heavy}$ relative to $\it {light}$ $s$-process elements. This is customarily
characterized by the ratio [hs/ls] (Luck \& Bond 1991) where we take La and Nd
as
heavy-$s$ (hs) and Y and Zr as light-$s$ (ls) elements. 
The abundances of these four $s$-process elements are represented
by several lines and available for all the C-rich pAGBs. Busso et al. (2001)
also used the same $s$-process elements in their theoretical computations.
The results would not be significantly changed if we adopted any other combinations
of heavy $s$-process elements. 
The ratios [hs/ls] for 10 pAGBs including the two pAGBs studied here are given in Table~7.

The ratio [hs/ls] varies with metallicity across the pAGB sample. This is clearly
shown in Figure~8. For comparison we show the model predictions from Busso et al. (2001) for
the case in which the mass of $^{13}$C in the He-shell is assigned a constant and plausible value
independent of metallicity. The chosen value designated as ST/1.5 for 
a model of 1.5 M$_{\odot}$ by Busso et al. 
offers a good fit to [hs/ls] measurements for
$s$-process enriched stars over a wide range of metallicity. The model predictions
offer a good fit to the observed trend set by the pAGBs.
Two pAGB stars (IRAS~07134+1005 and IRAS~02229+6208) deviate from this trend and
they can be matched well with ST/3 model predictions. The models also
account for the observed enhancements [ls/Fe] and [hs/Fe].

\subsection{Lithium}
At [Fe/H] $\sim$ $-$0.8, the maximum lithium abundance seen in
main sequence stars is $\log\epsilon$(Li) $\simeq$ 2.3 (Chen et al.
2001). At the first dredge-up, the surface lithium abundance
is reduced by a factor of about 50, i.e., $\log\epsilon$(Li) $\simeq$ 0.6
is predicted for red giants. Shetrone, Sneden, \& Pilachowski (1993)
found lithium abundances from 0.3 to $<$ $-$0.5 in  old disk giants with
[Fe/H] similar to that of our pAGBs and a low $^{12}$C/$^{13}$C ratio. Hence, the pAGB stars with
$\log\epsilon$(Li) $\simeq$ 2.0 are very Li-rich relative to
their putative progenitors.
Therefore, the Li-rich stars have either experienced lithium
production, or external replenishment of lithium (e.g., engulfed
a planet).

Production is possible by what is known familiarly as the Cameron-Fowler (1971)
$^{7}$Be transport mechanism in which $^{3}$He by $\alpha$-capture is converted
to $^{7}$Be and subsequently to $^{7}$Li by electron capture. The $^{3}$He supply
is partly primordial $^{3}$He and partly the product of incomplete running of the $pp$-chain
in the outer parts of the low mass main sequence star. A high temperature is needed for $\alpha$-capture
on $^{3}$He but a low temperature is needed for $^{7}$Li (and $^{7}$Be) to avoid
destruction by protons. Thus, lithium production is efficient
only in a convective region/envelope with a hot base and cool outer boundary. In
intermediate mass (4 to 7M${\odot}$) AGB stars, predictions
and observations show that lithium production
is possible. This has been most clearly demonstrated for the most luminous AGB stars
in the Magellanic Clouds (Smith \& Lambert 1989, 1990; Plez, 
Smith, \& Lambert 1993; Sackmann \& Boothroyd 1992). The great majority of these Li-rich
stars are O-rich, a not unexpected result because the temperatures required for
$\alpha$-capture by $^{3}$He also ignite the CN-cycle which will convert
carbon to nitrogen and convert C-rich material to N-rich material with a low $^{12}$C/$^{13}$C ($\sim$ 3)
ratio.
Attribution of lithium production to intermediate-mass luminous AGB
stars implies that the Li-rich pAGBs evolved from stars with masses
of about 4~$M_\odot$, which is unlikely, although perhaps not
impossible, for stars as metal-poor as [Fe/H] $\sim$  $-$0.8.

Examination of the reaction rates shows that the time scales ($\propto n_i/(dn_i/dt)$) for
destruction of $^{3}$He and $^{12}$C are similar over the likely range of temperature.
The rate constant for $^{3}$He($\alpha, \gamma$)$^{7}$Be is approximately an order of magnitude
greater than for $^{12}$C(p, $\gamma$) $^{13}$N (Angulo et al. 1999) but this difference is
offset by the ratio $\alpha/p \simeq 0.1 $. Then, $^{3}$He and $^{12}$C are destroyed at about the same
rate. Unfortunately, the amount of $^{3}$He that was destroyed to produce
the observed lithium abundance is unknown because the efficiency of the
conversion of $^{7}$Be to $^{7}$Li is not obtainable from the
observations. Certainly, the observed lithium abundance is a small fraction (perhaps, a 10$^{6}$th)
of the $^{3}$He reservoir predicted by standard models. Even at low efficiency, a small
fractional change in the $^{3}$He abundance results in the observed lithium abundance with a similarly
small fractional change in the $^{12}$C and, hence, the $^{14}$N abundance. Provided
that fresh $^{12}$C is added from thermal pulses occuring during
and after lithium production, stars can remain C-rich with a high $^{12}$C/$^{13}$C
ratio yet present a lithium enrichment. 

Lithium-rich giants are known at much lower luminosities (cf.
Charbonnel \& Balachandran 2000). Several proposals exist in
which the $^3$He reservoir is tapped to provide the $^7$Li. It
is unlikely that these stars are progenitors of the Li-rich
pAGBs. The lithium in the giants will most probably be destroyed as the stars
evolve to the and up the AGB. Certainly, conversion of these O-rich giants
to C-rich AGB (and pAGB) stars will remove the lithium unless it is
replenished from the $^3$He reservoir.

Replenishment of lithium is in principle possible when the AGB star
engulfs a planet or a brown dwarf. Siess \& Livio (1999) discuss accretion
by a 3 M${\odot}$ AGB star. These calculations suggest that surface lithium is enhanced
by a factor of a few, a factor barely sufficient to account for the
observed lithium abundances. Although the parameter space of the scenario was not fully explored and 
other choices of AGB star (e.g., lower mass and lower metallicity) and accreted object may
result in a higher lithium abundance, this scheme will have difficulty in accounting for the
fact that a lithium enrichment is very common among the C-rich pAGB stars. Five of the 10 stars
in Table~7 have been reported to show lithium enrichment.
Spectra of the other five stars have been examined for lithium but the 6707\AA\ line was
apparently absent. Assuming a detection limit of 5 m\AA\ for the line we obtain the
abundance limits shown in Table~7. All are consistent with either lithium enrichment
or the low lithium abundance predicted for an AGB star that does not manufacture lithium.

\section{Conclusions}
Our abundance analysis based on high-resolution optical spectra of
IRAS 05113+1347 and IRAS 22272+5435 shows that these stars classified
as pAGB stars have the surface composition expected of a
highly-evolved AGB star: carbon and nitrogen enrichments (relative
to their presumed initial abundances) are identified as products of the
third and first dredge-ups, respectively, and a strong $s$-process
enrichment is also attributable to the third dredge-up. An intriguing
lithium enrichment may reflect lithium production from $^3$He in
the AGB or pAGB star. The two stars analysed in this paper seem
typical of the sample (now, 11 stars in total analysed) of C-rich
pAGB stars.

If the transition from AGB to pAGB star was made without drastic
alterations of surface composition, the immediate progenitor of a
C-rich pAGB was a cool carbon star of the N-type. Bright cool carbon
stars analysed by Lambert et al. (1986) and Ohnaka, Tsuji, \& Aoki
(2000) have C,N, and O abundances quite similar to the pAGBs, and
are $s$-process enriched to similar levels too. However, the carbon
stars are of approximately solar metallicity (Lambert et al. 1986;
Abia et al. 2001) not metal-poor like the pAGBs.
Lithium provides a discordant note. The N-type stars are not
lithium rich. The lithium-rich carbon stars
are the J-type stars rich in $^{13}$C and with no $s$-process
enhancement.
There are metal-poor carbon stars
in the halo that are as metal-poor as the pAGBs: Kipper et al.
(1996) analysed 5 stars with [Fe/H] from $-$0.7 to $-$1.2, all with
considerable $s$-process enrichments. With one exception, the carbon
abundances are similar to those of the pAGBs. But with the
same exception, the nitrogen abundances are much less than those of the
pAGBs. In two of the stars Li was found to be
enhanced (log $\epsilon$(Li) $\geq$ 1.7). The noticeable difference is that the $^{12}$C/$^{13}$C ratio of
the carbon stars are much lower (5 to 10) than the limits set for the
pAGBs.
In summary, the immediate progenitors of the pAGBs have yet to
be identified. This is odd given that the pAGBs evolve rapidly and
probably more rapidly than their AGB progenitor. It is quite
possible that the progenitor is shrouded by dust.

\acknowledgments
We acknowledge the support of National Science Foundation (Grant AST-9618414) and the Robert A. Welch Foundation
of Houston, Texas. We thank Dr. Pandey and Dr. Allende Prieto for many helpful comments. This research has made use
of the Simbad database, operated at CDS, Strasbourg, Francs, and the NASA ADS service, USA.

\clearpage

\figcaption[]{ The spectra of the two pAGBs from 6770 to 6800\AA\ showing
s-process lines.\label{fig1a}}

\figcaption[]{ The spectra of the two pAGBs from 7955 to 7980\AA\ showing
photospheric molecular CN lines (solid vertical marks) 
in IRAS~05113+1347. Note the absence of these CN lines
in the spectrum of IRAS~22272+5435. Telluric spectrum (dotted line)
is also shown to indicate the presence of atmospheric lines \label{fig1b}}

\figcaption[]{ Determination of atmospheric parameters: $T_{\rm{eff}}$, log~$g$, $\xi_{t}$, and
[M/H] using standard excitation and ionization balance among Fe\,{\sc i} and Fe\,{\sc ii} lines. 
In the top panel, the lower excitation potential (LEP)-abundance relation is used to determine $T_{\rm {eff}}$.
In the middle panel, the reduced equivalent width (W$_{\lambda}$/$\lambda$)-abundance
relation is used to determine $\xi_{\rm {t}}$, and, in the bottom
panel, the abundances of Fe\,{\sc i} (crosses) versus Fe\,{\sc ii} (hexagons)
are used in determining log $g$. In all the panels,
solid line is the least-squares fit to the data points.
\label{fig3}}

\figcaption[]{ Spectra of the two pAGBs showing
the circumstellar molecular C$_{2}$ (marked by vertical dashed lines) which are
strikingly narrow compared to their counterparts in the photospheric spectrum.
\label{fig4}}

\figcaption[]{ Profiles of Li\,{\sc i}, Na D$_{1}$, and K\,{\sc i} resonance lines
(December 31, 1999 spectra) for IRAS~05113+1347. Note the
strong emission in the Na\,{\sc i} profile which is very well aligned
with the central peaks of both the Li (dotted line) and K (broken line) profiles. A 
predicted K\,{\sc i} profile (solid line)
based on the derived abundances is also shown.
\label {fig5}}

\figcaption[]{ Profiles of Li\,{\sc i} (dotted line), Na D$_{1}$ (solid line),  and K\,{\sc i} (broken line) 
resonance lines for IRAS~22272+5435. 
Interstellar and telluric lines are marked by  {\it {IS}} 
and $\oplus$, respectively.
\label {fig6}}

\figcaption[]{Spectrum syntheses of the Li\,{\sc i} line at 6707.8\AA. 
Spectrum syntheses (solid lines) with
varying Li abundances are compared with the
observed Li profiles (dashed lines). \label{fig7}}

\figcaption[]{ Determination of the carbon isotopic ratio $^{12}$C/$^{13}$C for IRAS~05113+1347.
The 8001 - 8010\AA\ region was synthesized for three different ratios of
$^{12}$C/$^{13}$C = 25, 15, 5 (solid lines) and compared with observed spectrum (dashed line).
Abundances of C, N, and other elements are kept constant. \label{8}}

\figcaption[]{ Ratios of light (Y,Zr) and heavy (La and Nd) $s$-process
abundance ratios ([hs/ls]) of the intrinsic pAGB stars are plotted against metallicity ([Fe/H]).
Note that [hs/ls] increases with decreasing metallicity.
The dashed line is the predicted [hs/ls] versus [Fe/H] curve
from Busso et al. (2001) for ST/1.5 of 1.5 M$_{\odot}$ model \label{9}}

\clearpage

\begin{figure}
\plotone{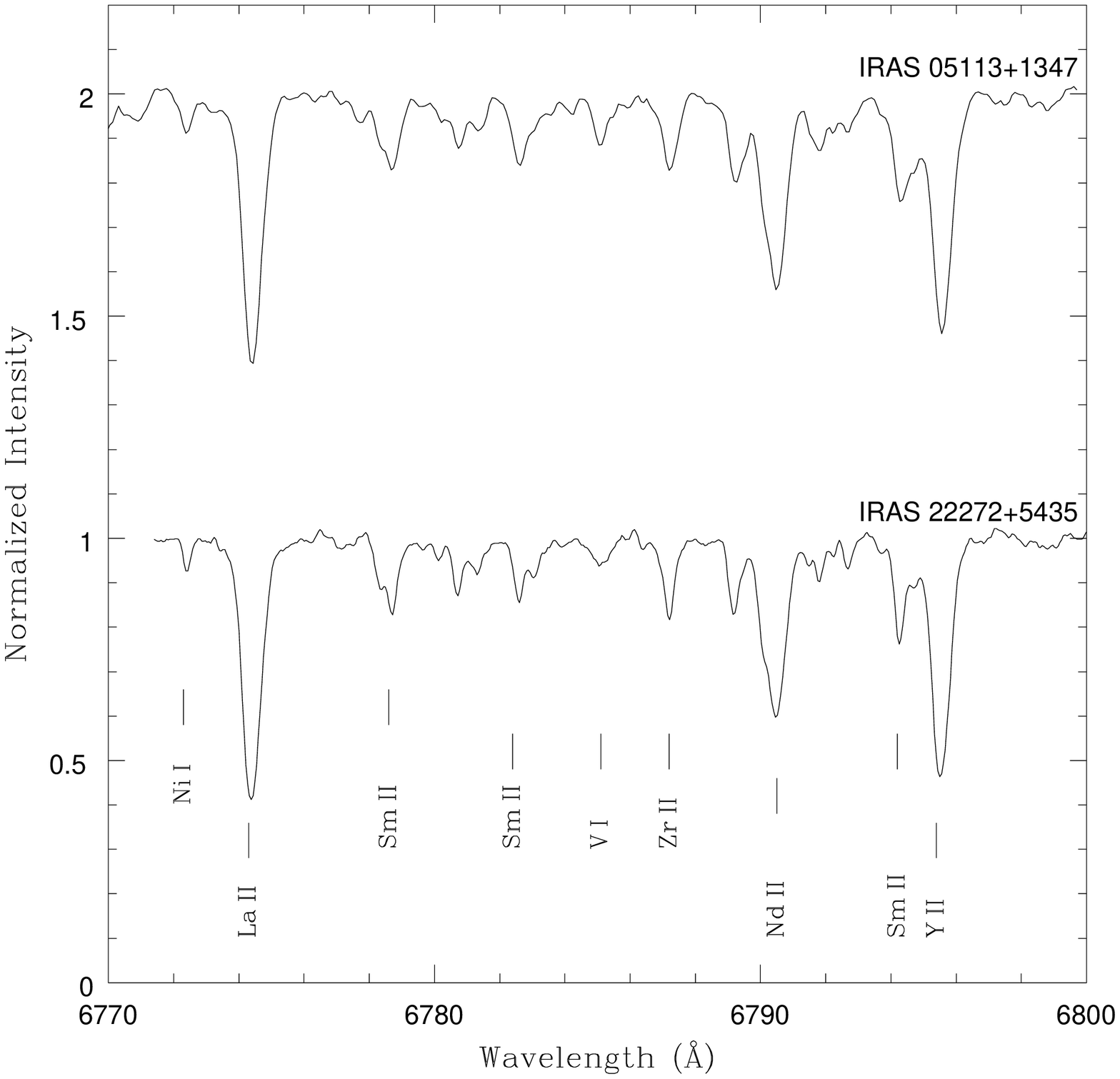}
\end{figure}

\clearpage

\begin{figure}
\plotone{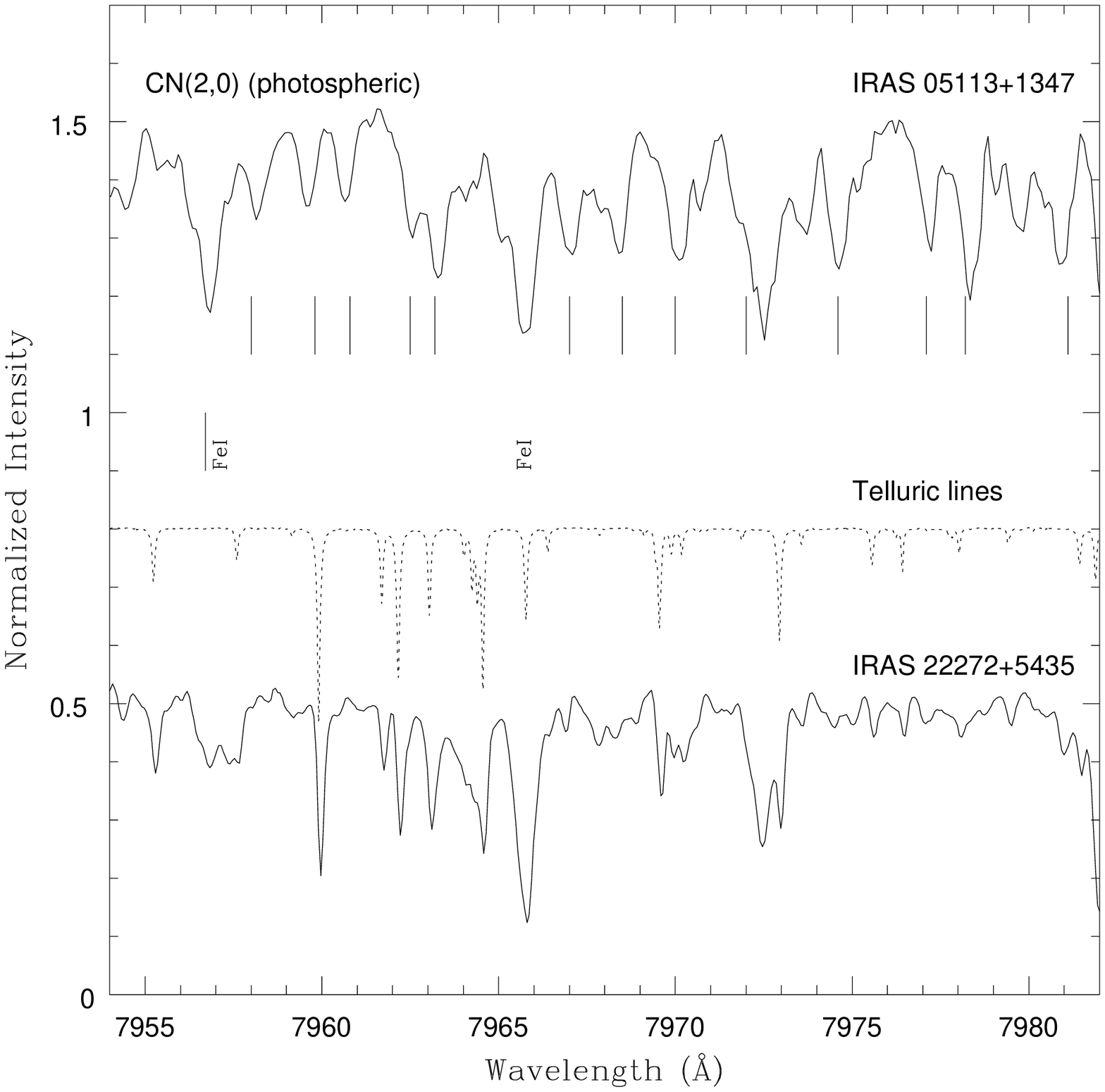}
\end{figure}

\clearpage

\begin{figure}
\plotone{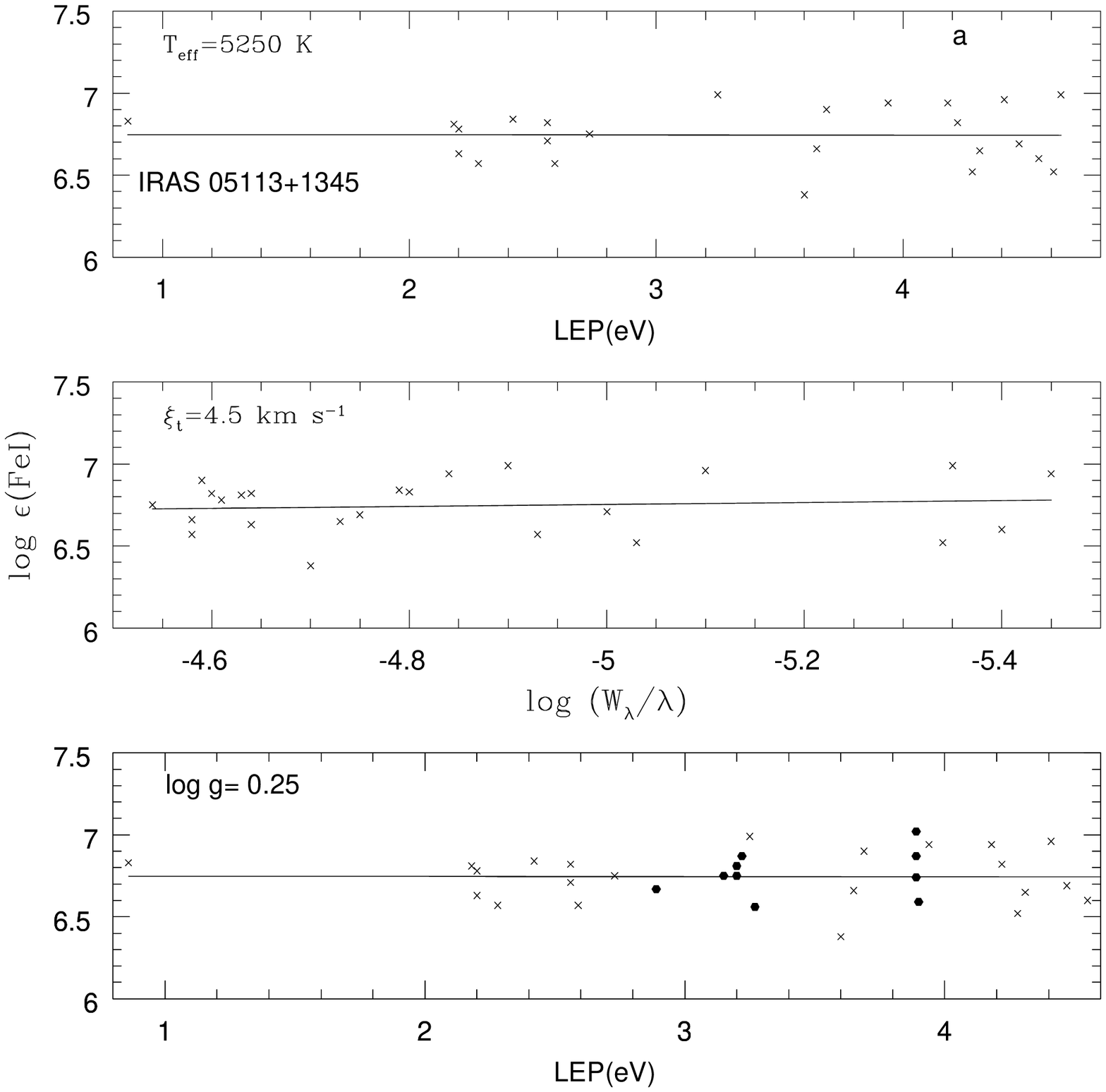}
\end{figure}

\clearpage

\begin{figure}
\plotone{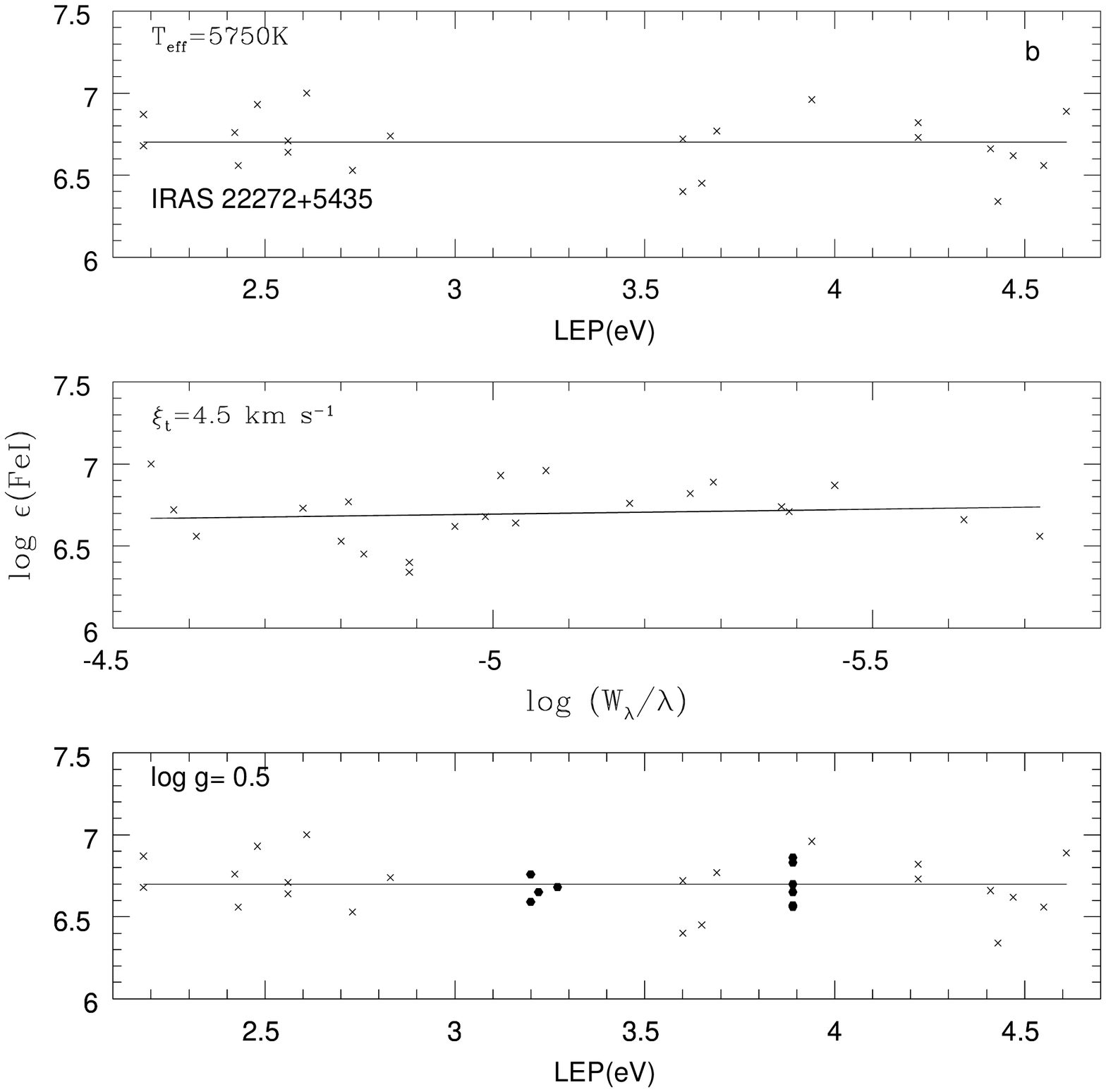}
\end{figure}

\clearpage

\begin{figure}
\plotone{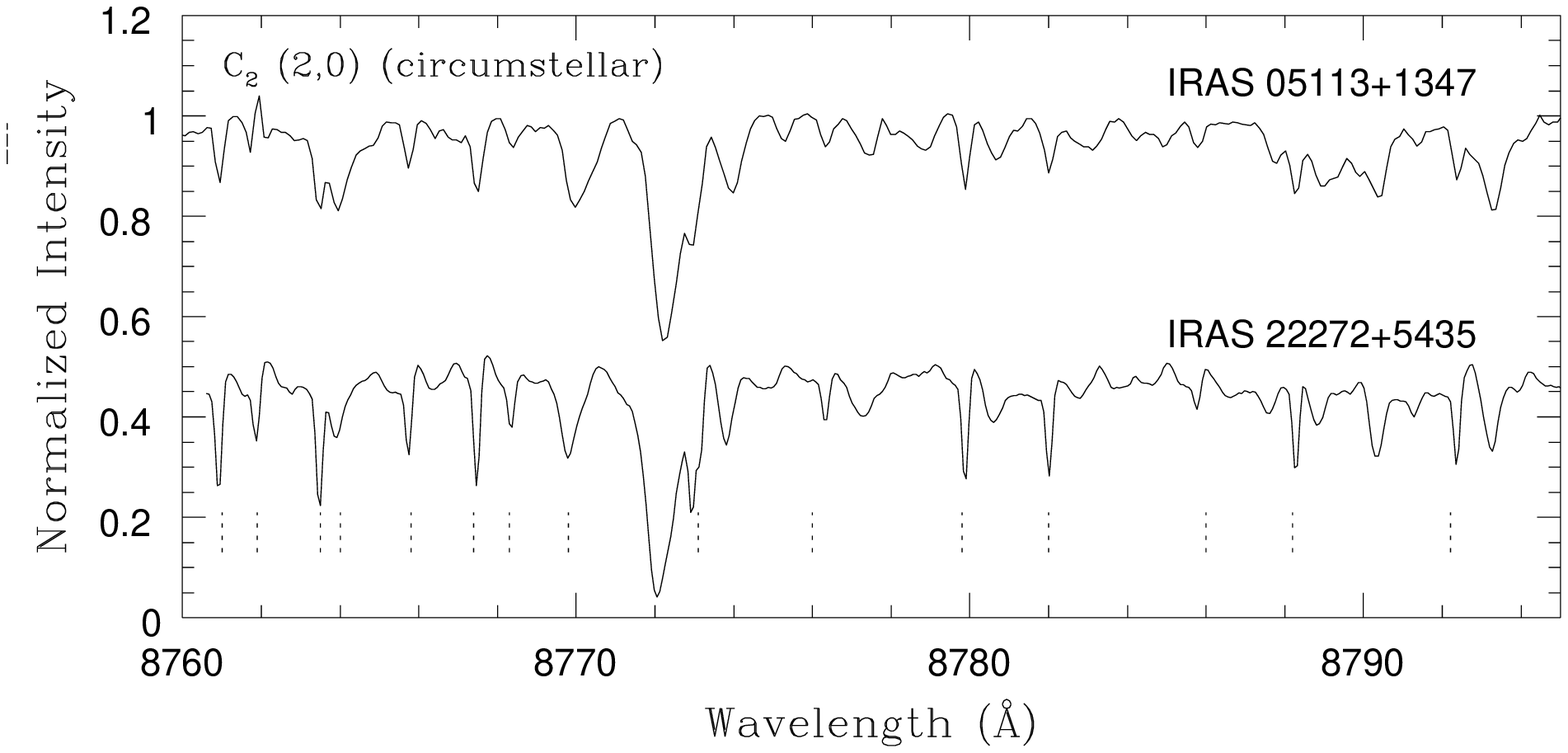}
\end{figure}

\clearpage

\begin{figure}
\plotone{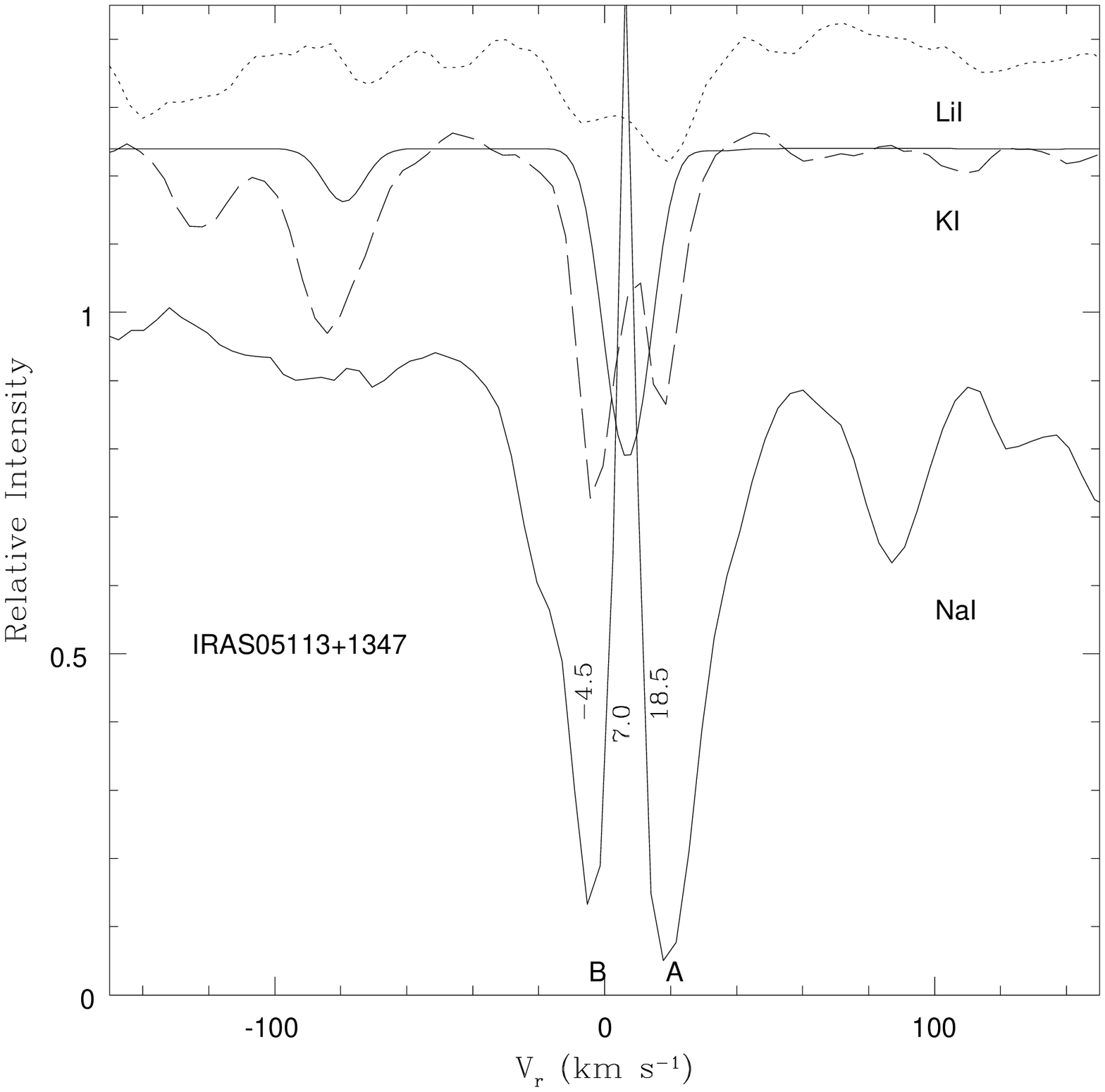}
\end{figure}

\clearpage
\begin{figure}
\plotone{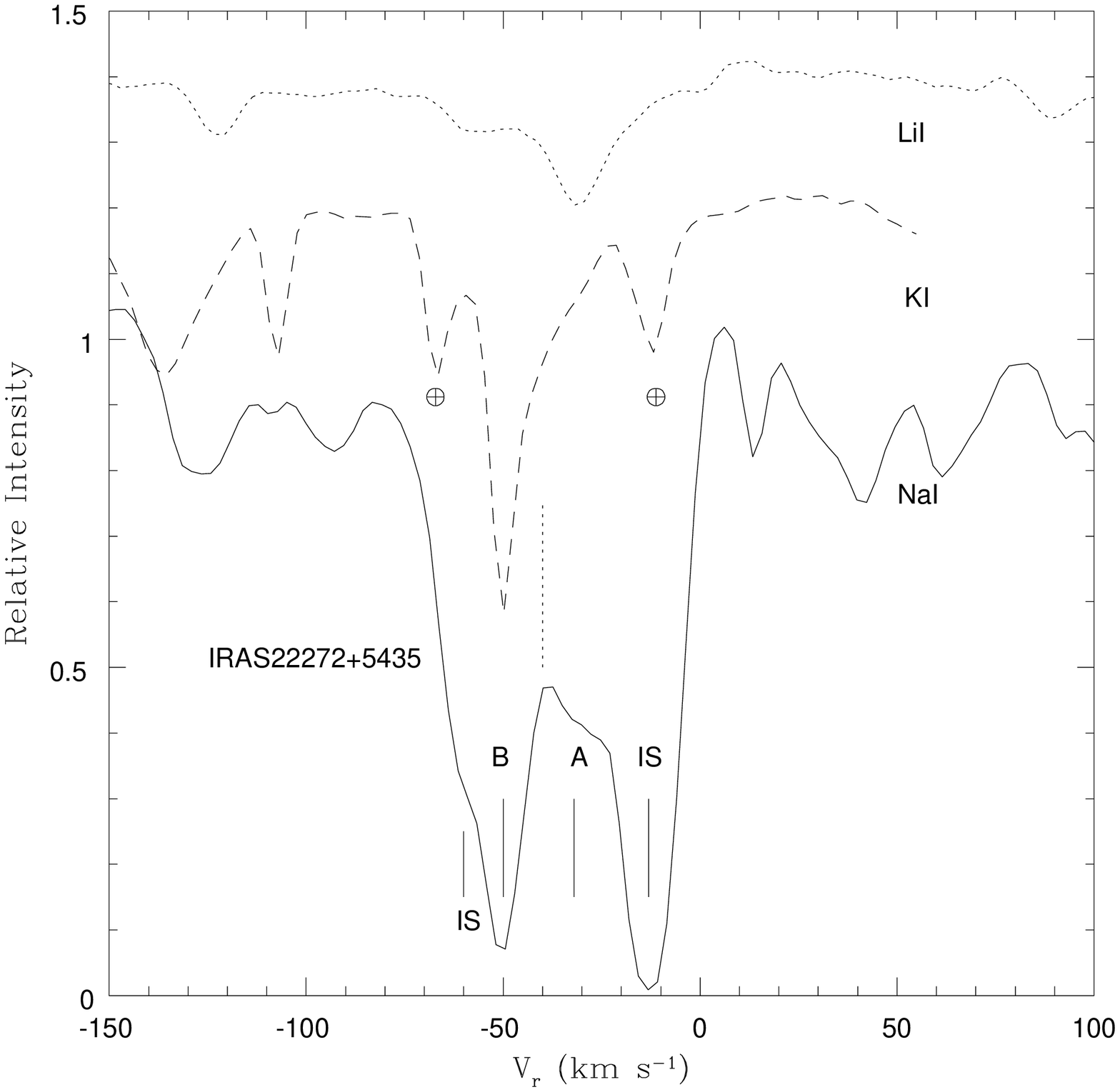}
\end{figure}

\clearpage

\begin{figure}
\plotone{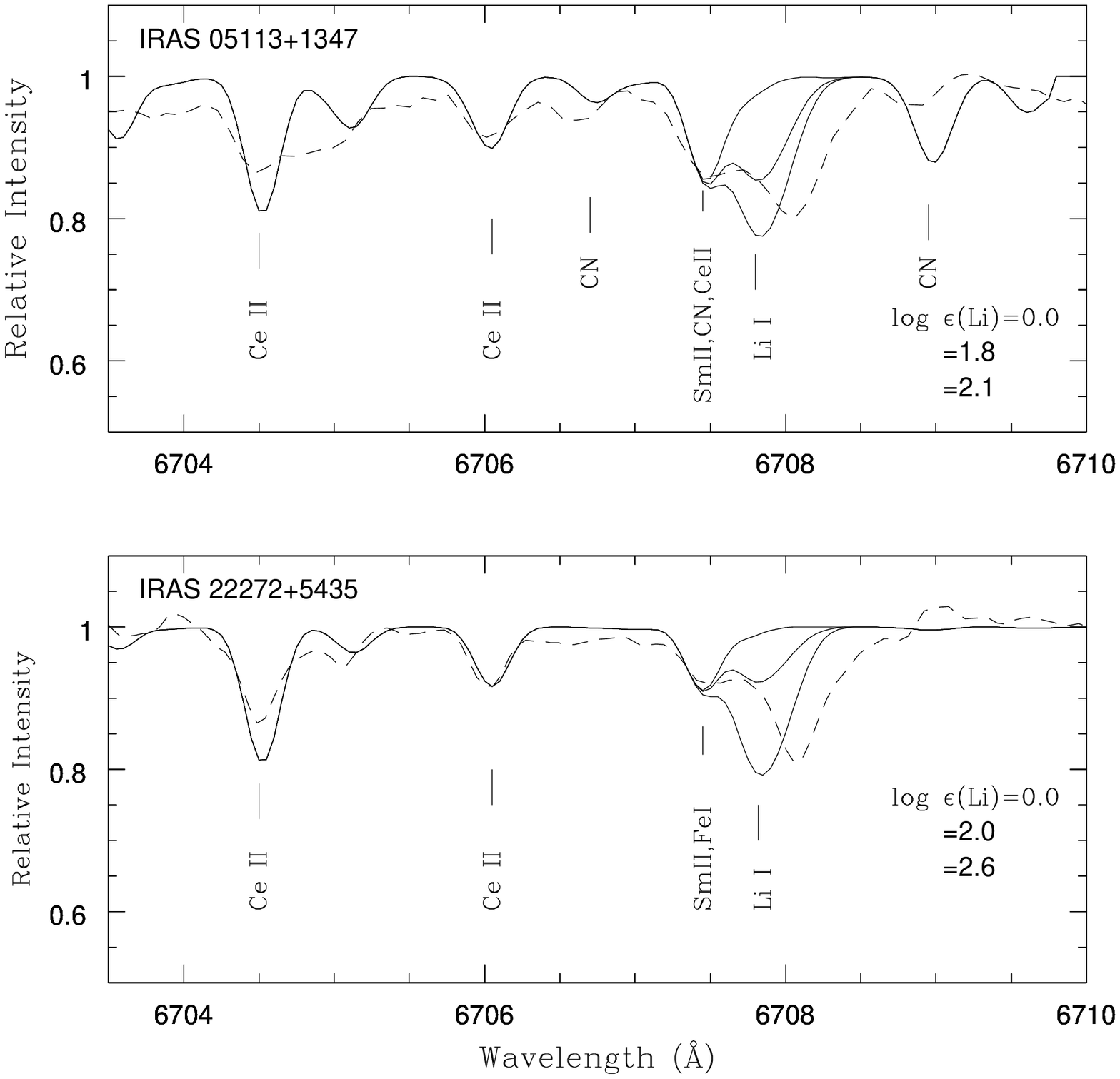}
\end{figure}

\clearpage

\begin{figure}
\plotone{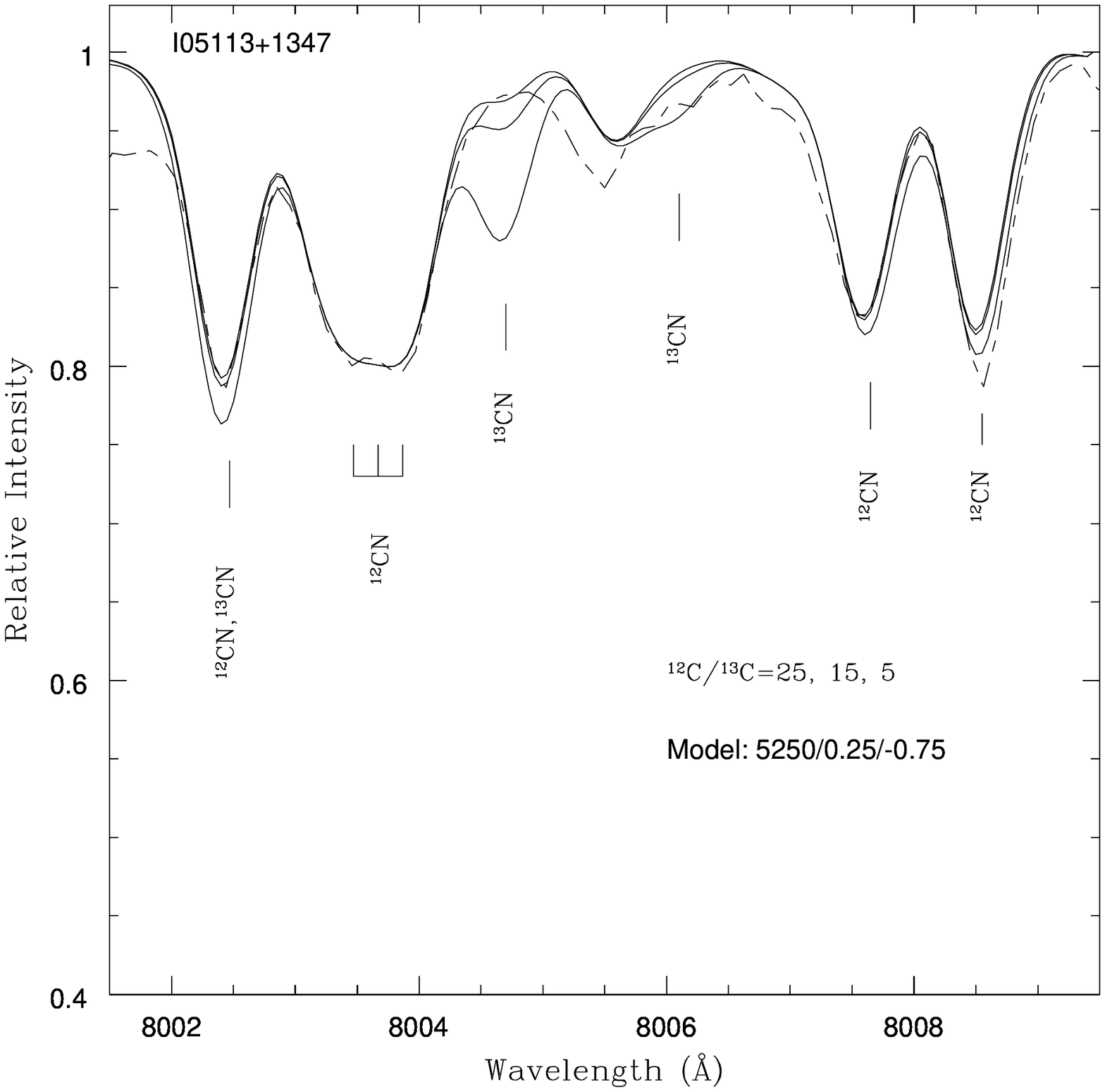}
\end{figure}

\clearpage

\begin{figure}
\plotone{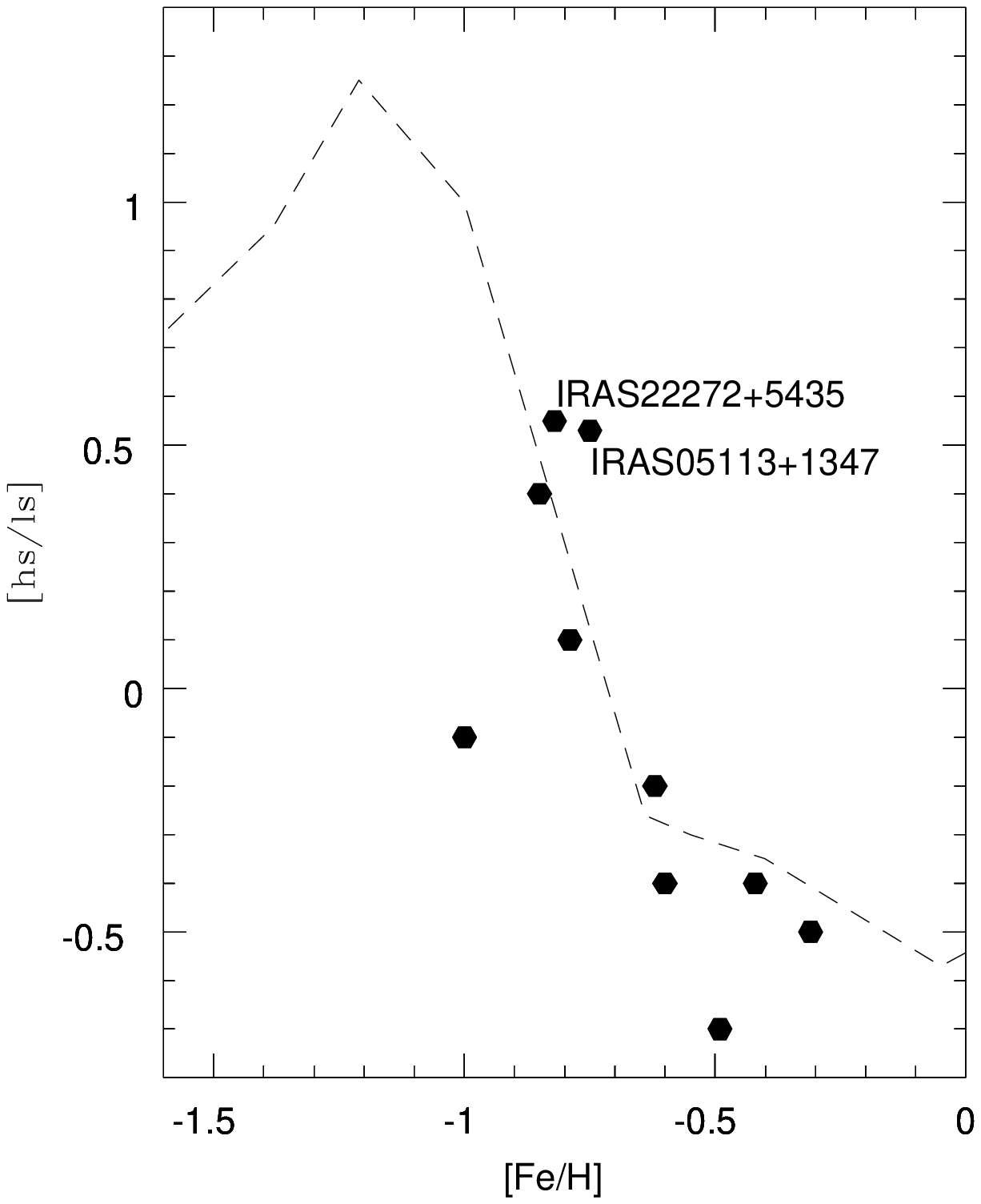}
\end{figure}

\clearpage

\begin{table}
\caption{Basic Observational Data }
\begin{tabular}{lcc}
\hline
\hline
      &  IRAS~05113+1347   & IRAS~22272+5435 \\
\hline
m$_{v}$    &    12.4            &  8.9 \\
B-V   &     2.1            &  1.9     \\
Sp.type &    G8Ia          &  G5Ia  \\
$l$   &    188.9           &  103.4      \\
$b$   &   $-$14.3          &  $-$2.5 \\
\hline
\end{tabular}
\end{table}

\begin{table}
\caption{Derived Spectroscopic Parameters}
\begin{tabular}{cccccccc}
\hline
\hline
IRAS       & $T_{\rm {eff}}$  &  log $g$ & $\xi_{t}$ & \lbrack {Fe/H}\rbrack & V$_{r}$ & V$_{exp}$\\     
           &   K            &    cm s$^{-2}$  &   km s$^{-1}$   &       &   km s$^{-1}$ &  km s$^{-1}$ \\
\hline        
05113+1347 &   5250   &  0.25 & 4.5 & $-$0.75 &   7.8  & 10.1 \\
22272+5435 &   5750   &  0.50 & 4.5 & $-$0.82 &  $-$42.3& 6.2\\ 
\hline
\end{tabular}
\end{table}
\clearpage

\begin{table}
\caption{Heliocentric radial velocity measurements for different species.
The expansion velocities (V$_{exp}$) are derived using molecular
C2 and CN lines which are originated in the dust envelope}
\begin{tabular}{llllllllll}
\hline
\hline
          &           & \multicolumn{3}{c}{IRAS~05113+1347}    & \multicolumn{3}{c}{IRAS~22272+5435} \\
\hline
Species   & LEP$^{a}$& $n$   & V$_{r}^{b}$  & $\sigma$ & $\it{n}$ & $V_{r}$  & $\sigma$ \\
\hline
C\,{\sc i}       &    8.8  & 4         &   5.0        &   1.0   &  8  & $-$43.4 & 1.9 \\
N\,{\sc i}       &    10.3 & 2        &  10.3        &  ...    &  2  & $-$43.3  & ...\\
S\,{\sc i}       &    8.0  & 3         &  8.6         &    2.5  &  2  & $-$41.1  & ...\\
Si\,{\sc i}      &    5.8  & 12         &  6.4         &    1.1 &  2  & $-$43.1  & ...\\
Fe\,{\sc i}     &     3.5  & 20         &  6.5         &    1.4  & 17 & $-$42.8& 1.2 \\
Fe\,{\sc ii}      &   3.6  & 11        &  7.8         &    1.1  & 33 & $-$41.6& 1.5 \\       
Ca\,{\sc i}      &     2.5 & 6         &  6.8         &    1.2  & 6  &$-$42.6  & 0.8  \\
Sc\,{\sc ii}     &     1.4 & 3         &  7.1         &    2.5  & 3  &$-$42.9  & 0.6  \\
Pr\,{\sc ii}     &     1.2 & 10         &  8.6         &    1.8  &  6 & $-$42.0  & 1.5  \\
Sm\,{\sc ii}     &     1.4 & 12         &  8.4         &    2.0  &  5 & $-$41.0  & 1.0 \\
\hline
\hline
\end{tabular}
\tablenotetext{a}{Average excitation potential (in eV) of the lower state of the transition}
\tablenotetext{b}{Mean radial velocity}

\end{table}

\clearpage

\begin{table}
\caption{ Abundance Results for pAGB stars: IRAs~05113$+$1347 and IRAS~22272+5435.
Abundances relative to both the solar ([X/H] and the iron ([X/Fe) are given.
The $\sigma$ denotes the scatter among the individual lines and $\it{n}$
represents number of lines used in the analysis.}
\begin{tabular}{cccccc|ccccc}
\hline
\hline
\multicolumn{5}{c}{IRAS~05113$+$1347}          & \multicolumn{4}{c} {IRAS~22272$+$5435} \\
Species  &  $\it {n}$  & log~$\epsilon$ (X) & \lbrack {X/H}\rbrack  & \lbrack {X/Fe} \rbrack & $\sigma$ 
         &  $\it {n}$  & log~$\epsilon$ (X) & \lbrack {X/H}\rbrack  & \lbrack {X/Fe} \rbrack & $\sigma$  \\
\hline
Li        & 1       &1.5 - 2.10&  -    & -     &  - &  1  &  2.0 - 2.5 & - & - & -\\ 
C\,{\sc i}       &   4     &    8.87     &   0.35   & 1.09 & 0.14   &  8  & 8.71 & 0.17  & 0.98  &  0.14 \\
\lbrack {C\,{\sc i}} \rbrack & 1 & 8.64     & 0.12  & 0.87 &  -  & -   &-  &  -    &   -   \\
N\,{\sc i}       &    2    &  8.23 & 0.32    & 1.05   & 0.11 &  2  &7.68 & -0.24 &  0.57 & 0.13 \\
O\,{\sc i}       &    1    &  8.54 &-0.29    & 0.45   &  - &  -  & - &  - & -    \\
\lbrack {O\,{\sc i}} \rbrack & 2 & 8.37 &  $-$0.46  & 0.29 & - & 1  & 8.53 &  -0.30    &  0.50    &   -   \\
Na\,{\sc i}      &   1     & 5.82 &  -0.51    & 0.24   &  -  &   2 & 5.89 & -0.44 &  0.30 &  0.04 \\
Al\,{\sc i}      &   3     & $<$ 5.72 & $<$ -0.75 & $<$ 0.0 & - &   2 & 5.78  & -0.69 &  0.13  & ... \\
Si\,{\sc i}      &   4     & 7.09 & -0.46    & 0.28 & 0.06  &   2  & 7.17  & -0.60 &  0.15  & 0.06 \\
S\,{\sc i}       &   5     & 6.87 &-0.52    & 0.23 & 0.22  &   6  & 6.89 & -0.60 &  0.15  & 0.20 \\
Ca\,{\sc i}      &   6     & 5.64 &-0.72    & 0.02 & 0.15  &   5  & 5.68 &-0.68 &  0.05  & 0.18 \\
Ti\,{\sc i}      &   5     & 4.60 & -0.42   & 0.32 & 0.09  &  2   &  4.55& -0.45  &  0.37 & - \\
Ti\,{\sc ii}     &   5     & 4.36 & -0.66   & 0.08 & 0.25  &  3   &  4.46 &  -0.54 &  0.28 & 0.11 \\
Cr\,{\sc i}      &   5     & 4.89 &-0.78   & -0.04 & 0.09 &  5 & 5.13   & -0.54 &  0.27  & 0.04 \\
Mn\,{\sc i}      &   3     & 4.68 &-0.78    & 0.04  & 0.05 &  2   &4.80 & -0.59 &  0.22  & 0.15 \\
Fe\,{\sc i}      &  24     & 6.74 & -0.76    & 0.00  & 0.17 & 22   &6.70 & -0.80 &  0.00  & 0.18 \\
Fe\,{\sc ii}     & 11      & 6.77 & -0.73    & 0.00  & 0.14 & 13   &6.67 & -0.83 &  0.00  & 0.10 \\
Ni\,{\sc ii}      & 12      & 5.45 &-0.80    & -0.06 & 0.16 & 6 &5.53    & -0.72 &  0.02  & 0.17 \\ 
Zn\,{\sc i}      &  1      & 3.74 &-0.85    & -0.10 &  -   & 1 & 4.16    & -0.58 &  0.23  &  -   \\
Y\,{\sc ii}      &  4      & 3.36 & 1.12    & 1.86  & 0.07 & 3 &3.28   & 1.04  & 1.81   & 0.14 \\ 
Zr\,{\sc ii}     &  2      & 3.23 & 0.63    & 1.37  & 0.00 & 2 &3.17   & 0.57  & 1.31   & 0.04 \\
La\,{\sc ii}     &  10     & 2.71 & 1.49    & 2.23  & 0.19 & 9  &2.75  &  1.53 & 2.27   & 0.18 \\
Ce\,{\sc ii}     &  12     & 2.84 & 1.30    & 2.04  & 0.20 & 6  & 2.76  &  1.22 & 2.03   & 0.14 \\
Pr\,{\sc ii}     &  10     & 1.63 & 0.92    & 2.56  & 0.20 & 6  & 1.61  & 0.90   & 1.71  & 0.21 \\
Nd\,{\sc ii}     &  8      & 2.61 & 1.11    &  1.85 & 0.05 & 7  & 2.76  & 1.30   & 2.11  & 0.10 \\
Sm\,{\sc ii}     &  12     & 2.36 & 1.36    & 2.10  & 0.10 & 5  & 2.37   & 1.90  & 2.71  & 0.17 \\
\hline
\end{tabular}
\end{table}

\clearpage

\begin{table}
\caption{Uncertainties in the Abundance ratios [X/Fe]
for the C-rich pAGB star IRAS~05113~1347, due to Uncertainties
in the Model Atmospheric Parameters}
\begin{tabular}{cccccccc}
\hline
\hline
\lbrack{x/Fe}\rbrack & LEP$^{a}$ & $\delta$ $T_{\rm{eff}}$  & $\delta$ $log g$ &$\delta$ $\xi_{t}$ 
& $\delta$ \lbrack{M/H}\rbrack & $\sigma_{m}$ \\  
                    &     Ev         &  $+$150~K   &   $+$0.50   & $+$0.5 km s$^{-1}$  & $+$0.25~dex &  \\
\hline
        Fe   &3.3& -0.09  &   -0.06  &   0.06 &   0.010  &    0.12 \\
        C\,{\sc i}  &8.6&  0.24  &   -0.13  &  -0.03 &   -0.01  &    0.27\\
       \lbrack{C\,{\sc i}}\rbrack &1.3&  0.05  &   -0.12  &  -0.03 &   -0.03  &    0.14\\
        N   &10.3&  0.30  &   -0.15  &  -0.04 &    0.02  &    0.34\\
        O   & 0.0&  0.09  &   -0.12  &  -0.03 &   -0.02  &    0.15\\
        Na  & 2.1 &-0.02  &    0.09  &  -0.06 &   -0.01  &    0.11\\
        Al  & 4.0 & 0.03  &    0.08  &  -0.02 &    0.01  &    0.09\\
        Si  & 5.8 & 0.02  &    0.09  &  -0.04 &    0.00  &    0.10\\
        S   & 8.1 & 0.18  &   -0.10  &  -0.04 &   -0.01  &    0.21\\
        Ca  & 2.6 &-0.05  &    0.10  &   0.00 &    0.01  &    0.11\\
        Sc  & 1.4 & 0.02  &   -0.12  &   0.05 &   -0.02  &    0.13\\
       Ti\,{\sc i} & 1.4 &-0.11  &    0.10  &  -0.04 &    0.01  &    0.15\\
       Ti\,{\sc i}I &2.0 & 0.05  &   -0.11  &  -0.02 &   -0.02  &    0.12\\
        V  &  0.8 &-0.12  &    0.11  &  -0.06 &    0.00  &    0.17\\
        Cr &  1.8 &-0.09  &    0.11  &  -0.02 &    0.00  &    0.14\\
        Mn &  1.7 &-0.12  &    0.11  &  -0.04 &    0.00  &    0.17\\
        Ni &  3.3 &-0.08  &    0.10  &  -0.02 &    0.01  &    0.13\\
        Zn &  3.7 &-0.06  &    0.09  &   0.02 &    0.01  &    0.11\\
        Y  &  1.7 & 0.02  &   -0.11  &   0.03 &   -0.02  &    0.12\\
        Zr &  2.5 & 0.05  &   -0.11  &  -0.03 &   -0.03  &    0.13\\
        La &  1.5 & 0.02  &   -0.11  &   0.02 &   -0.02  &    0.11\\
        Ce &  1.7 & 0.02  &   -0.11  &  -0.03 &   -0.04  &    0.12\\
        Pr &  1.2 &-0.03  &   -0.09  &  -0.01 &   -0.03  &    0.10\\
        Nd &  1.3 &-0.03  &   -0.10  &  -0.02 &   -0.03  &    0.11\\
        Sm &  1.3 &-0.02  &   -0.10  &  -0.04 &   -0.03  &    0.11 \\
        Eu &  1.3 & 0.00  &   -0.14  &   0.03 &   -0.02  &    0.15\\
\hline
\end{tabular}
\tablenotetext{a}{Average excitation potential (in eV) of the lower state of the transition}
\end{table}

\clearpage

\begin{table}
\caption{Abundances of C, N, and O.
Ratios are calculated after correcting
for initial C and O Abundances at the observed [Fe/H] values. }
\begin{tabular}{lllllllllll}
\hline
\hline
Star     & \lbrack Fe/H\rbrack       &\multicolumn{2}{c}{log $\epsilon$(C)}&
\multicolumn{2}{c}{log $\epsilon$(N)}&
\multicolumn{2}{c}{log $\epsilon$(O)} & C/O  & N/O & $\Sigma${CNO/Fe} \\
                &   & C$_{ini}$ & C$_{obs}$&   N$_{ini}$ & N$_{obs}$&   O$_{ini}$ & O$_{obs}$&  & \\
\hline
IRAS~05113+1347 &$-$0.74 & 7.97&  8.81 & 7.18 & 8.24  & 8.31& 8.43  & 2.57  & 1.07  & 0.59 \\
IRAS~22272+5435 &$-$0.82 & 7.90&  8.69 & 7.10 & 7.68  & 8.26& 8.48  & 1.82  & 0.28  & 0.51 \\
                &        &     &       &      &       &             &       &       &     \\
IRAS~02229+6208 &$-$0.45 & 8.21&  8.84 & 7.47 & 8.67  & 8.50& ...   & ...   & ...   & ...     \\
IRAS~04296+3429 &$-$0.70 & 8.00&  8.71 & 7.30 & 7.76  & 8.34& ...   & ...   & ...   & ...   \\
IRAS~05341+0852 &$-$0.85 & 7.88&  8.73 & 7.07 & 7.83  & 8.24& 8.57  & 1.62  & 0.33  & 0.60 \\
IRAS~07134+1005 &$-$1.00 & 7.76&  8.65 & 7.46 & 7.84  & 8.20& 8.67  & 1.15  & 0.31  & 0.71 \\
IRAS~07430+1115 &$-$0.46 & 8.20&  8.76 & 6.92 & 7.97  & 8.49& ...   & ...   & ...   & ...  \\
IRAS~19500-1709 &$-$0.60 & 8.09&  8.98 & 7.32 & 8.37  & 8.40& 8.96  & 1.05  & 0.38  & 0.75 \\
IRAS~22223+4327 &$-$0.46 & 8.20&  8.58 & 7.61 & 7.84  & 8.49& 8.50  & 1.15  & 0.29  & 0.20 \\
IRAS~23304+6147 &$-$0.79 & 7.93&  8.70 & 7.13 & 7.68  & 8.28& 8.24  & 3.16  & 0.48  & 0.43 \\
                &        &     &       &      &       &             &       &       &      \\
Mean Value      &$-$0.69 & 8.01&  8.75 & 7.26 & 7.99  & 8.35& 8.55  & 1.79  & 0.45  & 0.54  \\
Sun             &0.0 &... &  8.52 & ...  & 7.92  & ... & 8.83  & 0.49  & 0.12  & 0.0 \\
\hline
\end{tabular}
\end{table}

\clearpage

\begin{table}
\caption{ Results of Li, CNO, and carbon isotopic ratios for the two pAGB
stars are compared with the previously analyzed eight C-rich pAGB stars.
Upper limits to lithium abundances are based on assumed 
detectability limit of 5 m\AA\ for Li line at 6707\AA\ (See the text for the details).}
\begin{tabular}{llllllc}
\hline
\hline
pAGB    &  \lbrack Fe/H\rbrack  & log $\epsilon$ (Li)  & log $\epsilon$ (N) & $\Sigma{CN}_{initial}$&
$^{12}$C/$^{13}$C & \lbrack hs/ls\rbrack \\
\hline
IRAS~05113+1347$^{1}$ & $-$0.74        & 2.10  & 8.24 & 7.9 & $\geq$ 25$^{1}$ & 0.4  \\
                &                &       &      &      & $\geq$ 20$^{4}$  \\
IRAS~22272+5435$^{1}$ & $-$0.82        & 2.60  & 7.68 & 7.8 &  ...      & 0.6\\
IRAS~02229+6208$^{2}$ & $-$0.45  & 2.30  & 8.67 & 8.2 &  $\geq$ 25$^{1}$  & $-$0.9\\
IRAS~04296+3429$^{3}$ & $-$0.70  & $<$2.3  & 7.84 & 7.9 &  $\geq$20$^{4}$ & $-$0.3 \\
IRAS~05341+0852$^{2}$ & $-$0.85  & $\sim$2.5 & 7.83 & 7.8 & ...   & 0.6 \\
IRAS~07134+1005$^{3}$ & $-$1.00  & $<$2.7   & 7.84 & 7.6 &  ... & $-$0.1 \\
IRAS~07430+1115$^{2}$ & $-$0.46  & 2.40  & 7.97 & 8.2 &  ...  & $-$0.4\\
IRAS~19500-1709$^{3}$ & $-$0.60  & $<$3.6   & 8.37 & 8.2 &  ...  & $-$0.5 \\
IRAS~22223+4327$^{3}$ & $-$0.46  & $<$1.9   & 7.88 & 8.2 &  $\geq$20$^{4}$ & $-$0.4  \\
IRAS~23304+6147$^{3}$ & $-$0.79  & $<$2.1  &  7.68 & 7.8 & $\geq$20$^{4}$ & 0.2\\
\hline
\end{tabular}
\tablerefs{
(1) This work; (2) Reddy et al. (1996, 1999); (3) van Winckel $\&$ Reyniers (2000);
(4) Bakker et al. (1996)}
\end{table}
\clearpage

\begin{table}
\caption{Atomic Data and Abundances of Individual lines of $s$-process Elements}
\begin{tabular}{ccccccc}
\hline
\hline
           &      &           &\multicolumn{2}{c}{\underline{IRAS~05113+1347}} 
& \multicolumn{2}{c}{\underline{IRAS~22272+5435}}\\        
$\lambda$  & LEP  &           &$W_{\rm {\lambda}}$ &    &  $W_{\rm {\lambda}}$   \\
 (\AA)     & (eV) &  log $gf$ &(m\AA)             & log~$\epsilon$(X) & (\AA) & log~$\epsilon$(X) \\
\hline
{\underline{Y~II}}       &      &           &                   &     &           & \\
6832.47 & 1.74 & -1.940 & 188& 3.36   &  120 & 3.12  \\
6951.68 & 1.84 & -2.040 & 164& 3.36   &  ... & ...    \\
7332.96 & 1.72 & -2.480 & 113& 3.27   &  88  & 3.36   \\
6858.29 & 1.74 & -1.970 & 196& 3.44   &  150 & 3.36   \\
{\underline{Zr~II}}      &      &           &                   &     &           & \\
6578.65 & 2.43  & -1.500 &  68.0& 3.22  &  40  & 3.14\\ 
6787.16 & 2.49  & -1.170 & 102.0& 3.23  &  72  & 3.19 \\
{\underline{La~II}}      &      &           &                   &     &           & \\
6067.12  &    0.77  &  -2.340 &  130 &  2.58 & 92  &  2.66  \\ 
6126.07  &    1.25  &  -1.240 &  215 &  2.68 & 157 &  2.54 \\ 
6146.52  &    0.24  &  -2.470 &  180 &  2.46 & 143 &  2.61\\
6296.08  &    1.25  &  -0.950 &  261 &  2.76 & ... &  ...\\
6399.03  &    2.64  &  -0.530 &  117 &  2.71 & 85  &  2.67\\
6446.64  &    2.76  &  -0.720 &   81 &  2.74 & 75  &  2.89\\
6570.94  &    0.77  &  -2.520 &  142 &  2.79 & 118 &  3.00\\
6642.76  &    2.53  &  -0.960 &  122 &  3.04 & 65  &  2.79\\
6714.11  &    2.76  &  -0.930 &   78 &  2.91 & 65  &  2.99 \\
6834.10  &    0.24  &  -2.180 &  219 &  2.39 & 185 &  2.57 \\
{\underline{Ce~II}}      &      &           &                   &     &           & \\
7105.05 &     0.54  &  -2.526  &   110.0  &    3.04  &  ...  & ... \\
7486.56 &     2.55  &  -0.394  &    52.0  &    2.63  &  ...  & ... \\
7689.18 &     1.58  &  -1.165  &   115.0  &    2.84  &  88   & 2.98 \\
7313.45 &     1.93  &  -0.919  &   123.0  &    3.07  &  54   & 2.81 \\  
6706.05 &     1.84  &  -1.253  &    26.0  &    2.40  &  24   & 2.66 \\  
6652.74 &     1.53  &  -1.213  &   152.0  &    3.17  & ...   &  ... \\   
6393.03 &     1.46  &  -1.242  &    96.0  &    2.73  & 64    &  2.80 \\   
6466.89 &     1.77  &  -1.022  &   104.0  &    2.91  & 50    &  2.75 \\   
6973.50 &     2.32  &  -0.518  &    55.0  &    2.56  & 33    &  2.55 \\   
7058.69 &     0.29  &  -2.761  &   129.0  &    3.13  & ...   &   ... \\   
7061.75 &     1.93  &  -0.640  &   148.0  &    2.99  & ...   &   ... \\   
7086.35 &     2.13  &  -0.441  &    98.0  &    2.65  & ...   &   ... \\   
{\underline{Pr~II}}      &      &           &                   &     &           & \\
6656.83 &     1.82  &   0.082  &   131.0  &   1.74  &  66 & 1.58 \\
6566.76 &     0.22  &  -1.721  &   133.0  &    1.76 &  ... & ... \\
6429.63 &     1.61  &  -0.326  &    68.0  &    1.45 & ...  & ... \\
\hline
\end{tabular}
\end{table}

\clearpage
\begin{table}
\begin{tabular}{ccccccc}
\multicolumn{7}{l}{Table~8-Continued}\\
\hline
\hline
           &      &           &\multicolumn{2}{c}{\underline{IRAS~05113+1347}} 
& \multicolumn{2}{c}{\underline{IRAS~22272+5435}}\\        
$\lambda$  & LEP  &           &$W_{\rm {\lambda}}$ &    &  $W_{\rm {\lambda}}$   \\
 (\AA)     & (eV) &  log $gf$ &(m\AA)             & log~$\epsilon$(X) & (\AA) & log~$\epsilon$(X) \\
\hline
{\underline{Pr~II}}      &      &           &                   &     &           & \\
6363.63 &     1.12  &  -1.419  &    35.0  &    1.64 & 28  & 1.92 \\
6413.68 &     1.13  &  -0.853  &    ...   &    ...  & 36  & 1.49 \\ 
6281.28 &     0.96  &  -0.570  &   117.0  &    1.36 & ...  & ... \\
6165.89 &     0.92  &  -0.205  &   199.0  &    1.59 & 166  & 1.74 \\
6093.06 &     1.46  &  -0.648  &    39.0  &    1.32 & ...  & ... \\
6087.53 &     1.12  &  -0.618  &   124.0  &    1.66 & ...  & ...  \\ 
6086.18 &     1.30  &  -0.801  &    92.0  &    1.81 & ...  &  ... \\
6016.51 &     1.00  &  -0.798  &   152.0  &    1.92 & ...  & ... \\
6182.34 &     1.44  &  -0.387  &   ...    &    ...  & 33.0  & 1.31 \\ 
6046.65 &     1.12  &  -0.686  &   ...    &    ...  & 58.0  & 1.60 \\
{\underline{Nd~II}}      &      &           &                   &     &           & \\
6248.27 &     1.22 &   -1.600  &    93.0  &    2.56 & 60 & 2.67 \\
6250.44 &     1.16 &   -1.620  &   112.0  &    2.66 & 79 & 2.80 \\
6591.43 &     0.20 &   -2.510  &   126.0  &    2.54 & 76 & 2.66 \\
6637.19 &     1.45 &   -1.080  &   143.0  &    2.63 & 125 & 2.88 \\
6650.52 &     1.95 &   -0.170  &   184.0  &    2.60 & 147 & 2.64 \\ 
6669.63 &     1.04 &   -1.720  &   115.0  &    2.61 & 83  & 2.78 \\
6680.14 &     1.69 &   -0.810  &   148.0  &    2.66 & ... &  ... \\
6698.64 &     1.64 &   -1.360  &    80.0  &    2.65 & 63  & 2.86 \\
{\underline{Sm~II}}      &      &           &                   &     &           & \\
7085.49 &     1.07 &   -1.963  &   100.0  &     2.52  &  ... & ... \\
6844.70 &     1.36 &   -1.529  &    95.0  &     2.38  &  ... & ... \\
6734.05 &     1.37 &   -1.400  &   106.0  &     2.37  &  83  & 2.53 \\
6731.81 &     1.17 &   -1.372  &   145.0  &     2.38  &  117 & 2.55 \\
6694.72 &     1.37 &   -2.209  &    30.0  &     2.43  &   ... & ... \\
6693.56 &     1.69 &   -1.089  &   104.0  &     2.39  &   ... & ... \\
6679.22 &     1.07 &   -1.691  &   108.0  &     2.34  &   ... & ... \\
6632.27 &     1.67 &   -1.344  &    53.0  &     2.19  &   38 & 2.33 \\
6616.61 &     1.35 &   -2.220  &    30.0  &     2.42  &   ... & ... \\
6544.61 &     1.17 &   -2.180  &    30.0  &     2.18  &   13 & 2.14 \\
6542.76 &     1.17 &   -1.737  &    77.0  &     2.26  &   45 & 2.31 \\
6484.55 &     1.26 &   -1.792  &    81.0  &     2.46  &   ... & ... \\
\hline
\end{tabular}
\end{table}

\end{document}